\input harvmac
\input epsf
\noblackbox

\newcount\figno

\figno=0
\def\fig#1#2#3{
\par\begingroup\parindent=0pt\leftskip=1cm\rightskip=1cm\parindent=0pt
\baselineskip=11pt \global\advance\figno by 1 \midinsert
\epsfxsize=#3 \centerline{\epsfbox{#2}} \vskip 12pt
\centerline{{\bf Figure \the\figno :}{\it ~~ #1}}\par
\endinsert\endgroup\par}
\def\figlabel#1{\xdef#1{\the\figno}}
\def\pano{\par\noindent}

\font\cmss=cmss10
\font\cmsss=cmss10 at 7pt

\def\rlx{\relax\leavevmode}
\def\inbar{\vrule height1.5ex width.4pt depth0pt}
\def\IC{\relax\,\hbox{$\inbar\kern-.3em{\rm C}$}}
\def\IR{\relax{\rm I\kern-.18em R}}
\def\IN{\relax{\rm I\kern-.18em N}}
\def\IP{\relax{\rm I\kern-.18em P}}
\def\frac#1#2{{#1 \over #2}}
\def\ZZ{\rlx\leavevmode\ifmmode\mathchoice{\hbox{\cmss Z\kern-.4em Z}}
 {\hbox{\cmss Z\kern-.4em Z}}{\lower.9pt\hbox{\cmsss Z\kern-.36em Z}}
 {\lower1.2pt\hbox{\cmsss Z\kern-.36em Z}}\else{\cmss Z\kern-.4em Z}\fi}

\def\narrowplus{\kern -.04truein + \kern -.03truein}
\def\narrowminus{- \kern -.04truein}
\def\narrowminussub{\kern -.02truein - \kern -.01truein}

\def\a{\alpha}

\def\o#1{\overline{#1}}

\def\ra{\rangle}



\lref\SchellekensAM{
A.~N.~Schellekens and S.~Yankielowicz,
``Extended Chiral Algebras And Modular Invariant Partition Functions,''
Nucl.\ Phys.\ B {\bf 327}, 673 (1989)\semi
A.~N.~Schellekens and S.~Yankielowicz,
``Modular Invariants From Simple Currents: An Explicit Proof,''
Phys.\ Lett.\ B {\bf 227}, 387 (1989)\semi
A.~N.~Schellekens and S.~Yankielowicz,
``New Modular Invariants For N=2 Tensor Products And Four-Dimensional
Strings,''
Nucl.\ Phys.\ B {\bf 330}, 103 (1990)\semi
K.~A.~Intriligator,
``Bonus Symmetry In Conformal Field Theory,''
Nucl.\ Phys.\ B {\bf 332}, 541 (1990).
}

\lref\GreeneUD{
B.~R.~Greene and M.~R.~Plesser,
``Duality In Calabi-Yau Moduli Space,''
Nucl.\ Phys.\ B {\bf 338}, 15 (1990).
}

\lref\rkakufour{Z.~Kakushadze,
``Aspects of N = 1 type I-heterotic duality in four dimensions,''
Nucl.\ Phys.\ B {\bf 512}, 221 (1998)
[arXiv:hep-th/9704059]\semi
Z.~Kakushadze and G.~Shiu,
``A chiral N = 1 type I vacuum in four dimensions and its heterotic dual,''
Phys.\ Rev.\ D {\bf 56}, 3686 (1997)
[arXiv:hep-th/9705163].
}

\lref\rang{C.~Angelantonj, M.~Bianchi, G.~Pradisi, A.~Sagnotti and Y.~S.~Stanev,
``Chiral asymmetry in four-dimensional open- string vacua,''
Phys.\ Lett.\ B {\bf 385}, 96 (1996)
[arXiv:hep-th/9606169].
}

\lref\riban{G.~Aldazabal, A.~Font, L.~E.~Ibanez and G.~Violero,
``D = 4, N = 1, type IIB orientifolds,''
Nucl.\ Phys.\ B {\bf 536}, 29 (1998)
[arXiv:hep-th/9804026]\semi
G.~Aldazabal, L.~E.~Ibanez, F.~Quevedo and A.~M.~Uranga,``D-branes at singularities: A bottom-up approach to the string  embedding of
JHEP {\bf 0008}, 002 (2000)
[arXiv:hep-th/0005067].
}

\lref\brunnew{I. Brunner, K. Hori, K. Hosomichi and J. Walcher,
``Orientifolds of Gepner Models,''
hep-th/0401137.
}

\lref\BlumenhagenSU{
R.~Blumenhagen,
``Supersymmetric orientifolds of Gepner models,''
JHEP {\bf 0311}, 055 (2003)
[arXiv:hep-th/0310244].
}

\lref\GepnerVZ{
D.~Gepner,
``Exactly Solvable String Compactifications On Manifolds Of SU(N) Holonomy,''
Phys.\ Lett.\ B {\bf 199}, 380 (1987).
}

\lref\GepnerQI{
D.~Gepner,
``Space-Time Supersymmetry In Compactified String Theory And Superconformal Models,''
Nucl.\ Phys.\ B {\bf 296}, 757 (1988).
}

\lref\IntriligatorZW{
K.~A.~Intriligator,
``Bonus Symmetry In Conformal Field Theory,''
Nucl.\ Phys.\ B {\bf 332}, 541 (1990).
}
\lref\IntriligatorZW{
K.~A.~Intriligator,
``Bonus Symmetry In Conformal Field Theory,''
Nucl.\ Phys.\ B {\bf 332}, 541 (1990).
}

\lref\UrangaPZ{
A.~M.~Uranga,
``Chiral four-dimensional string compactifications with intersecting  D-branes,''
Class.\ Quant.\ Grav.\  {\bf 20}, S373 (2003)
[arXiv:hep-th/0301032].
}

\lref\RecknagelSB{
A.~Recknagel and V.~Schomerus,
``D-branes in Gepner models,''
Nucl.\ Phys.\ B {\bf 531}, 185 (1998)
[arXiv:hep-th/9712186].
}

\lref\review{
A.~M.~Uranga,
``Chiral four-dimensional string compactifications with intersecting  D-branes,''
Class.\ Quant.\ Grav.\  {\bf 20}, S373 (2003), [arXiv:hep-th/0301032]\semi
}

\lref\BrunnerNK{
M.~Gutperle and Y.~Satoh,
``D-branes in Gepner models and supersymmetry,''
Nucl.\ Phys.\ B {\bf 543}, 73 (1999)
[arXiv:hep-th/9808080]\semi
S.~Govindarajan, T.~Jayaraman and T.~Sarkar,
``Worldsheet approaches to D-branes on supersymmetric cycles,''
Nucl.\ Phys.\ B {\bf 580}, 519 (2000)
[arXiv:hep-th/9907131]\semi
D.~E.~Diaconescu and C.~R\"omelsberger,
``D-branes and bundles on elliptic fibrations,''
Nucl.\ Phys.\ B {\bf 574}, 245 (2000)
[arXiv:hep-th/9910172]\semi
P.~Kaste, W.~Lerche, C.~A.~Lutken and J.~Walcher,
``D-branes on K3-fibrations,''
Nucl.\ Phys.\ B {\bf 582}, 203 (2000)
[arXiv:hep-th/9912147]\semi
E.~Scheidegger,
``D-branes on some one- and two-parameter Calabi-Yau hypersurfaces,''
JHEP {\bf 0004}, 003 (2000)
[arXiv:hep-th/9912188]\semi
M.~Naka and M.~Nozaki,
``Boundary states in Gepner models,''
JHEP {\bf 0005}, 027 (2000)
[arXiv:hep-th/0001037].
}

\lref\brunschom{
I.~Brunner and V.~Schomerus,
``D-branes at singular curves of Calabi-Yau compactifications,''
JHEP {\bf 0004}, 020 (2000)
[arXiv:hep-th/0001132].
}

\lref\BrunnerJQ{
I.~Brunner, M.~R.~Douglas, A.~E.~Lawrence and C.~R\"omelsberger,
``D-branes on the quintic,''
JHEP {\bf 0008}, 015 (2000)
[arXiv:hep-th/9906200].
}

\lref\MizoguchiXI{
S.~Mizoguchi and T.~Tani,
``Wound D-branes in Gepner models,''
Nucl.\ Phys.\ B {\bf 611}, 253 (2001)
[arXiv:hep-th/0105174].
}

\lref\BrunnerZM{
I.~Brunner and K.~Hori,
``Orientifolds and mirror symmetry,''
arXiv:hep-th/0303135.
}

\lref\BrunnerEM{
I.~Brunner and K.~Hori,
``Notes on orientifolds of rational conformal field theories,''
[arXiv:hep-th/0208141].
}

\lref\BrunnerFS{
I.~Brunner,
``On orientifolds of WZW models and their relation to geometry,''
JHEP {\bf 0201}, 007 (2002)
[arXiv:hep-th/0110219].
}

\lref\HuiszoonAI{
L.~R.~Huiszoon and K.~Schalm,
``BPS orientifold planes from crosscap states in Calabi-Yau  compactifications,''
JHEP {\bf 0311}, 019 (2003) [arXiv:hep-th/0306091].
}

\lref\HuiszoonVY{
L.~R.~Huiszoon,
``Comments on the classification of orientifolds,''
Class.\ Quant.\ Grav.\  {\bf 20}, S509 (2003)
[arXiv:hep-th/0212244].
}

\lref\FuchsCM{
J.~Fuchs, L.~R.~Huiszoon, A.~N.~Schellekens, C.~Schweigert and J.~Walcher,
``Boundaries, crosscaps and simple currents,''
Phys.\ Lett.\ B {\bf 495}, 427 (2000)
[arXiv:hep-th/0007174].
}

\lref\HuiszoonGE{
L.~R.~Huiszoon and A.~N.~Schellekens,
``Crosscaps, boundaries and T-duality,''
Nucl.\ Phys.\ B {\bf 584}, 705 (2000)
[arXiv:hep-th/0004100].
}

\lref\HuiszoonXQ{
L.~R.~Huiszoon, A.~N.~Schellekens and N.~Sousa,
``Klein bottles and simple currents,''
Phys.\ Lett.\ B {\bf 470}, 95 (1999)
[arXiv:hep-th/9909114].
}

\lref\GovindarajanVP{
S.~Govindarajan and J.~Majumder,
``Crosscaps in Gepner models and type IIA orientifolds,''
[arXiv:hep-th/0306257].
}

\lref\GovindarajanVV{
S.~Govindarajan and J.~Majumder,
``Orientifolds of type IIA strings on Calabi-Yau manifolds,''
arXiv:hep-th/0305108.
}

\lref\BlumenhagenJY{
R.~Blumenhagen, D.~L\"ust and S.~Stieberger,
``Gauge unification in supersymmetric intersecting brane worlds,''
JHEP {\bf 0307}, 036 (2003)
[arXiv:hep-th/0305146].
}

\lref\BlumenhagenTJ{
R.~Blumenhagen and A.~Wisskirchen,
``Spectra of 4D, N = 1 type I string vacua on non-toroidal CY threefolds,''
Phys.\ Lett.\ B {\bf 438}, 52 (1998)
[arXiv:hep-th/9806131].
}

\lref\AngelantonjMW{
C.~Angelantonj, M.~Bianchi, G.~Pradisi, A.~Sagnotti and Y.~S.~Stanev,
``Comments on Gepner models and type I vacua in string theory,''
Phys.\ Lett.\ B {\bf 387}, 743 (1996)
[arXiv:hep-th/9607229].
}

\lref\AldazabalUB{
G.~Aldazabal, E.~C.~Andres, M.~Leston and C.~Nunez,
``Type IIB orientifolds on Gepner points,''
JHEP {\bf 0309}, 067 (2003)
[arXiv:hep-th/0307183].
}

\lref\FuchsGV{
J.~Fuchs, C.~Schweigert and J.~Walcher,
``Projections in string theory and boundary states for Gepner models,''
Nucl.\ Phys.\ B {\bf 588}, 110 (2000)
[arXiv:hep-th/0003298].
}

\lref\RecknagelQQ{
A.~Recknagel,
``Permutation branes,''
JHEP {\bf 0304}, 041 (2003)
[arXiv:hep-th/0208119].
}

\lref\stanev{
Y.~S.~Stanev,''Open Descendants of Gepner Models in 6D'', talk presented at the
workshop ``Conformal Field Theory of D=branes'', at DESY, Hamburg, Germany,
Transparancies on http://www.desy.de/$\sim$jfuchs/CftD.html.
}

\lref\rbgklnon{R.~Blumenhagen, L.~G\"orlich, B.~K\"ors and D.~L\"ust,
``Noncommutative Compactifications of Type I Strings on Tori with Magnetic
Background Flux'', JHEP {\bf 0010}, 006 (2000), [arXiv:hep-th/0007024] \semi
R.~Blumenhagen, B.~K\"ors and D.~L\"ust,
``Type I Strings with $F$ and $B$-Flux'', JHEP {\bf 0102}, 030 (2001),
[arXiv:hep-th/0012156].
}

\lref\raads{C.~Angelantonj, I.~Antoniadis, E.~Dudas, A.~Sagnotti, ``Type I
Strings on Magnetized Orbifolds and Brane Transmutation'',
Phys. Lett. B {\bf 489}, 223 (2000), [arXiv:hep-th/0007090].
}

\lref\timo{R.~Blumenhagen and T.~Weigand. {\it in preparation}.}

\lref\rbbkl{R.~Blumenhagen, V.~Braun, B.~K\"ors and D.~L\"ust,
``Orientifolds of K3 and Calabi-Yau Manifolds with Intersecting D-branes'',
JHEP {\bf 0207}, 026  (2002), [arXiv:hep-th/0206038].
}

\lref\rcveticb{M.~Cvetic, G.~Shiu and  A.M.~Uranga,  
``Chiral Four-Dimensional N=1 Supersymmetric Type IIA Orientifolds from
Intersecting D6-Branes'', Nucl. Phys. B {\bf 615}, 3  (2001), [arXiv:hep-th/0107166]\semi
R.~Blumenhagen, L.~G\"orlich and T.~Ott,
``Supersymmetric intersecting branes on the type IIA 
$T^6/Z(4)$  orientifold'', JHEP {\bf 0301}, 021 (2003),
[arXiv:hep-th/0211059]\semi
G.~Honecker,
``Chiral supersymmetric models on an orientifold of Z(4) x Z(2) with 
 intersecting D6-branes,''
Nucl.\ Phys.\ B {\bf 666}, 175 (2003)
[arXiv:hep-th/0303015]\semi
M.~Larosa and G.~Pradisi,
``Magnetized four-dimensional Z(2) x Z(2) orientifolds,''
Nucl.\ Phys.\ B {\bf 667}, 261 (2003)
[arXiv:hep-th/0305224].
}

\lref\rbachas{C.~Bachas, ``A Way to Break Supersymmetry'', [arXiv:hep-th/9503030]\semi 
M.~Berkooz, M.R.~Douglas and R.G.~Leigh, ``Branes Intersecting
at Angles'', Nucl. Phys. B {\bf 480}, 265  (1996), [arXiv:hep-th/9606139].
}

\lref\rafiru{G.~Aldazabal, S.~Franco, L.E.~Ibanez, R.~Rabadan, A.M.~Uranga,
{\it $D=4$ Chiral String Compactifications from Intersecting Branes},
J.\ Math.\ Phys.\  {\bf 42}, 3103 (2001), [arXiv:hep-th/0011073]\semi
G.~Aldazabal, S.~Franco, L.E.~Ibanez, R.~Rabadan, A.M.~Uranga,
{\it Intersecting Brane Worlds}, JHEP {\bf 0102}, 047 (2001), [arXiv:hep-ph/0011132].
}

\lref\BianchiYU{
M.~Bianchi and A.~Sagnotti,
``On The Systematics Of Open String Theories,''
Phys.\ Lett.\ B {\bf 247}, 517 (1990)\semi
M.~Bianchi and A.~Sagnotti,
``Twist Symmetry And Open String Wilson Lines,''
Nucl.\ Phys.\ B {\bf 361}, 519 (1991)\semi
M.~Bianchi, G.~Pradisi and A.~Sagnotti,
``Toroidal compactification and symmetry breaking in open string theories,''
Nucl.\ Phys.\ B {\bf 376}, 365 (1992).
}

\lref\PradisiQY{
G.~Pradisi, A.~Sagnotti and Y.~S.~Stanev,
``Planar duality in SU(2) WZW models,''
Phys.\ Lett.\ B {\bf 354}, 279 (1995)
[arXiv:hep-th/9503207]\semi
G.~Pradisi, A.~Sagnotti and Y.~S.~Stanev,
``The Open descendants of nondiagonal SU(2) WZW models,''
Phys.\ Lett.\ B {\bf 356}, 230 (1995)
[arXiv:hep-th/9506014]\semi
G.~Pradisi, A.~Sagnotti and Y.~S.~Stanev,
``Completeness Conditions for Boundary Operators in 2D Conformal Field Theory,''
Phys.\ Lett.\ B {\bf 381}, 97 (1996)
[arXiv:hep-th/9603097].
}

\lref\CallanPX{
C.~G.~Callan, C.~Lovelace, C.~R.~Nappi and S.~A.~Yost,
``Adding Holes And Crosscaps To The Superstring,''
Nucl.\ Phys.\ B {\bf 293}, 83 (1987)\semi
J.~Polchinski and Y.~Cai,
``Consistency Of Open Superstring Theories,''
Nucl.\ Phys.\ B {\bf 296}, 91 (1988).

}

\lref\DouglasSW{
M.~R.~Douglas and G.~W.~Moore,
``D-branes, Quivers, and ALE Instantons,''
arXiv:hep-th/9603167.
}

\Title{\vbox{
 \hbox{DAMTP-2004-2}
 \hbox{hep-th/0401148}}}
{\vbox{\centerline{ Chiral Supersymmetric} 
\vskip 0.3cm \centerline{Gepner Model Orientifolds}
}}
\centerline{Ralph Blumenhagen{$^1$}, Timo Weigand{$^2$} }
\bigskip\medskip
\centerline{ {\it DAMTP, Centre for Mathematical Sciences,}}
\centerline{\it Wilberforce Road, Cambridge CB3 0WA, UK}
\centerline{\tt Email:\vtop{{\hbox{$^1$: R.Blumenhagen@damtp.cam.ac.uk}}
                            {\hbox{$^2$: T.Weigand@damtp.cam.ac.uk}}} }
\bigskip
\bigskip

\centerline{\bf Abstract}
\noindent
We explicitly construct A-type orientifolds of supersymmetric
Gepner models. In order to reduce the  tadpole
cancellation conditions to a treatable number we explicitly work out 
the generic form  of  the one-loop Klein bottle, annulus
and M\"obius strip amplitudes
for simple current extensions of Gepner models. 
Equipped with these formulas, we discuss two examples
in detail to provide evidence that in this setting certain
features of the MSSM  like unitary gauge groups with large enough rank, 
chirality and family replication can be achieved.


\Date{01/2004}
\newsec{Introduction}
 
Various ways for constructing phenomenologically semi-realistic  four-dimensional
string vacua
have been followed in the past. After focusing solely on the heterotic
string, the realization that D-branes play a very crucial role
in string theory opened up the possibility
to look out for new string vacua where these D-branes
are actually present. The right framework to study such models
are orientifold models, where the presence of the D-branes
is implied by tadpole cancellation conditions. 
These orientifold models were the subject of intense study
during the last years and it became clear that there
are essentially two approaches to get chiral semi-realistic models. 
One can either get chiral fermions from D-branes on orbifold
singularities \refs{\DouglasSW\rang\rkakufour-\riban}
or from intersecting D-branes 
\refs{\rbachas\rbgklnon\raads\rafiru\rcveticb-\UrangaPZ}. Both approaches
have been followed extensively and many semi-realistic models
have been proposed so far, but it is certainly fair so say that
none of them gives indeed rise to physics of the Standard Model, but
at least some of its rough data could be realized such as
the right gauge group and chiral fermions in three generations.  

So far these constructions were mostly limited to toroidal orbifold
backgrounds, which, of course, cover only  a very small
subset of all possible closed string backgrounds. 
It is therefore desirable to also study orientifold
models on more general non -flat Calabi-Yau spaces \rbbkl. One way
to approach this problem is to start with a class of  exactly
solvable superconformal field theories which are known
to describe certain points deep inside the K\"ahler 
 moduli space of Calabi-Yau manifolds. These models
are commonly known as Gepner models  \refs{\GepnerQI,\GepnerVZ}.
The construction of boundary 
\refs{\RecknagelSB\brunschom\BrunnerNK\BrunnerJQ\FuchsGV\MizoguchiXI-\RecknagelQQ}
and crosscap states
 \refs{\BianchiYU\PradisiQY\HuiszoonXQ\HuiszoonGE\FuchsCM\BrunnerFS
\HuiszoonVY\BrunnerEM\BrunnerZM\GovindarajanVV\HuiszoonAI-\GovindarajanVP}
in these models has been under investigation during the last years. 
However, there have been only quite a few attempts so far to
really construct fully fledged orientifolds of Gepner models
\refs{\AngelantonjMW\stanev\BlumenhagenTJ\AldazabalUB-\BlumenhagenSU}.
Whereas in the beginning the construction of these orientifolds was carried
out in a case by case study, only very recently  \BlumenhagenSU\ it was possible 
to derive (at least for all levels being odd) 
quite generic formulas for all one-loop partition functions including
the highly non-trivial M\"obius strip amplitude containing
important sign factors. Moreover, amazingly simple expressions
for the tadpole cancellation conditions were derived. 
We regard these general equations as a useful starting point for
an explicit and general study of such models.
One has to distinguish two different kinds of orientifolds which are called
A-type and B-type. It turned out that for B-type models
the number of tadpole cancellation conditions is much smaller than for
the A-type models and therefore much easier to solve. 

This latter point was the main reason why in \BlumenhagenSU\ the primary
focus was on B-type models. Even though the tadpole cancellation
conditions could easily be solved, the prize one had to pay was that
due to general arguments these B-type models were always
non-chiral. Therefore from the phenomenological point of view
they are not very interesting. Additionally, from the technical point of view
it would have been desirable to confirm that the signs
in the M\"obius strip amplitude indeed lead to anomaly cancellation
of the effective low energy theory. But, of course, without
chiral fermions such a check could not be made. 

In this paper we continue the study of orientifolds of Gepner models
focusing on the A-type models, for which one expects that chiral models
should be possible. As we mentioned already, the main obstacle
is the huge number of tadpole cancellation conditions (of the order $10^2$),
which made it hard to determine any other solution than the generic
one with gauge group $SO(4)$ or $SP(4)$. 
In order to reduce this number we consider simple current
extensions \SchellekensAM\ of Gepner models and generalize the explicit
results for the one-loop amplitudes of 
\BlumenhagenSU\ to this case. Note that generic expressions for boundary
and crosscap states for simple current extensions 
were presented in \FuchsCM, which could also serve as an alternative 
starting point for such a construction.  For the set of simple
currents of Gepner models with all levels being odd we derive
the generic form of the Klein bottle, annulus and M\"obius
strip amplitudes. Equipped with these formulas, we study two models
in great detail and show that indeed certain aspects of the Standard Model,
including chirality, can be realized in this setting.
The aim of this paper is not to provide  a systematic search for
the MSSM in this class of models, but merely to provide
the relevant formalism and to give evidence that such a search
might be worthwhile carrying out.

This paper is organized as follows. In section 2, we review some
facts about Gepner models. In section 3, after some comments
about the simple current extension of modular invariants for general
CFTs, we apply these methods to the diagonal and charge conjugate
partition function of the Gepner model. Section 4 contains the
derivation of the corresponding orientifolds in terms of the computation 
of the simple current extended Klein bottle amplitude of A-type.
In principle due to the Greene-Plesser construction \GreeneUD\
of the mirror, A-type models would be sufficient, but
nevertheless we provide explicit formulas for the
B-type Klein bottle amplitude in Appendix A. 
In section 5, we review some of the important aspects of 
RS-boundary states including the loop and tree channel
annulus amplitudes. The derivation of the M\"obius strip
amplitudes is the subject of section 6, where for simplicity
we restrict ourselves to the explicit computation of the 
NS sector amplitudes. In section 7, we present the general tadpole
cancellation conditions, followed by a general analysis of the
massless open string sector including the gauge sector.
The techniques developed are employed to discuss a couple of examples
in section 8. Finally, section 9 contains our conclusions.

{\it Note:} While this work was in its very final stage  we became aware of the paper
Brunner et. al. \brunnew, which has some overlap with our work. 

\newsec{Review of Gepner models}

Let us  briefly review some aspects of Gepner models
needed in the following. 
In light-cone gauge, the internal sector of a Type II compactification
to four dimensions with N=2 supersymmetry is given
by tensor products of the rational models of the N=2 super Virasoro
algebra with total central charge $c=9$ \refs{\GepnerQI,\GepnerVZ}. 
 Space-time supersymmetry is achieved by  a GSO projection, which can be described
by a certain simple current in the superconformal field theory.

 The minimal models are parametrized by the 
level $k=1,2,\ldots$ and have central charge
\eqn\num{
c={3k\over k+2} .}
Since $c<3$, one achieves the required value $c=9$ by using  tensor products of 
such minimal models
$ \bigotimes_{j=1}^r (k_j)$. 
The finite number
of irreducible representations of the N=2 Virasoro algebra of each
unitary model 
are labeled by the three integers $(l,m,s)$ in the range
\eqn\range{  l=0,\ldots k, \quad m=-k-1,-k,\ldots k+2, \quad
              s=-1,0,1,2 }
with $l+m+s=0$ mod $2$. 
Actually, the identification between $(l,m,s)$ and
$(k-l,m+k+2,s+2)$ reveals that the range \range\  is a double covering of the
allowed representations. 
The conformal dimension and charge of the highest weight 
state with label $(l,m,s)$ is given by
\eqn\quannum{\eqalign{
\Delta^l_{m,s}&={l(l+2)-m^2\over 4(k+2)} + {s^2\over 8} \cr
q^l_{m,s}&={m\over (k+2)}-{s\over 2}  .}}
Note that these formulas are  only correct modulo one and two respectively. 
To obtain the precise conformal dimension $h$ and charge  from \quannum\
one  first shifts  the labels into the standard range $|m-s|\le l$
by using the shift symmetries $m\to m+2k+4$,$s\to s+4$ and 
the reflection symmetry.  
The NS-sector consists of those representations with even $s$, while
the ones
with odd $s$ make up to the R-sector.

In addition to the internal N=2 sector, one has the contributions with $c=3$ from the 
two uncompactified
directions. The two world-sheet fermions $\psi^{2,3}$ 
generate a $U(1)_2$ model whose four irreducible
representations are labeled by $s_0=-1,\ldots 2$ with highest weight
and charge modulo one and two respectively
\eqn\fermi{ \Delta_{s_0}={s_0^2\over 8}  \ ,
             \quad\quad q_{s_0}=-{s_0\over 2} .}
The GSO projection means in the Gepner case that one projects onto states 
with odd overall $U(1)$ charge $Q_{tot}=q_{s_0}+\sum_{j=1}^r q^{l_j}_{m_j,s_j}$.
Moreover, to have a good space-time interpretation one has to ensure
that in the tensor product only states from the NS respectively 
the R sectors couple among themselves.
 
These projections are described most conveniently in the following notation.
First one defines some multi-labels
\eqn\mlabels{  \lambda=(l_1,\ldots,l_r), \quad \mu=(s_0;m_1,\ldots m_r; 
                s_1,\ldots,s_r) }
and the respective characters 
\eqn\mchar{ \chi^\lambda_\mu(q)=\chi_{s_0}(q)\, \chi^{l_1}_{m_1,s_1}(q)
                        \ldots \chi^{l_r}_{m_r,s_r}(q) .}
In terms of the vectors 
\eqn\mvec{ \beta_0=(1;1,\ldots,1;1,\ldots,1), \quad 
            \beta_j=(2;0,\ldots,0;0,\ldots,0,\underbrace{2}_{j^{\rm th}},0,\ldots,0) }
and the following product
\eqn\mprod{\eqalign{  Q_{tot}&=2\beta_0\bullet \mu =-{s_0\over 2}-\sum_{j=1}^r{s_j\over 2}
              +\sum_{j=1}^r {m_j\over k_j+2},\cr
                  \beta_j\bullet \mu &=-{s_0\over 2}-{s_j\over 2},   }}
the projections one has to implement are simply 
$Q_{tot}=2\beta_0\bullet \mu\in 2\ZZ+1$ and $\beta_j\bullet \mu \in \ZZ$
for all $j=1,\ldots r$.
Gepner has shown that the following GSO projected
partition function
\eqn\parti{    Z_D(\tau,\o{\tau} )={1\over 2^r} 
{ ({\rm Im}\tau)^{-2} \over |\eta(q)|^4 }
     \sum_{b_0=0}^{K-1} \sum_{b_1,\ldots,b_r=0}^1 {\sum_{\lambda,\mu}}^\beta
    (-1)^{s_0} \ \chi^\lambda_\mu (q)\, \chi^\lambda_{\mu+b_0\beta_0
              +b_1 \beta_1 +\ldots b_r \, \beta_r} (\o q) }
is indeed modular invariant and vanishes due to space-time supersymmetry. 
Here $K={\rm lcm}(4,2k_j+4)$ and ${\sum}^\beta$ means that the sum is 
restricted to 
those $\lambda$ and $\mu$ in the range \range\ satisfying 
$2\beta_0\bullet \mu\in 2\ZZ+1$ and $\beta_j\bullet \mu \in \ZZ$.
The factor $2^r$ due to  the field
identifications guarantees the correct normalization of the amplitude. 
In the partition function \parti\ states with odd charge are arranged
in orbits under the action of the $\beta$ vectors. Therefore, although the partition
function is non-diagonal in the original characters, 
for all levels odd it can be written as a diagonal partition
function in terms of the GSO-orbits under the $\beta$-vectors \mvec, which in this case have
all equal length $2^r\,K$.

Let us also state the modular S-transformation
rules for the characters involved in \parti.
For the $SU(2)_k$
Kac-Moody algebra  the S-matrix  is given by
\eqn\ssmatrix{  S_{l,l'}=\sqrt{2\over k+2}\, \sin (l,l')_k ,}
where we have used the convention $(l,l')_k={\pi(l+1)(l'+1)\over k+2}$.
For the N=2 minimal model,  the modular S-matrix reads 
\eqn\smatrix{\eqalign{    S^{U(1)_2}_{s_0,s_0'}&={1\over 2} e^{-i\pi{s_0 s'_0
        \over 2}}, \cr
        {\cal S}_{(l,m,s),(l',m',s')}&={1\over 2\sqrt{2k+4}}\, S_{l,l'}\,
        e^{i\pi{m\, m' \over k+2}}\, e^{-i\pi{s\, s' \over 2}}  .}}
For a discussion of the normalization see \BlumenhagenSU.
  
The loop- and tree-channel M\"obius amplitudes are related by the P-matrix 
$P=T^{1\over 2}S\,T^2\,S\,T^{1\over 2}$, which for
 just the $SU(2)_k$ Kac-Moody algebra is given by
\eqn\ppmatrix{  P_{l,l'}={2\over \sqrt{k+2}}\, 
     \sin {1\over 2}(l,l')_k\, \delta_{l+l'+k,0}^{(2)} }
and for the N=2 unitary models  reads \BrunnerZM
\eqn\pmatrix{\eqalign{    P^{U(1)_2}_{s_0,s'_0}&={1\over \sqrt{2}}\, \sigma_{s_0}\sigma_{s'_0}
  e^{-i\pi{s_0 s'_0 \over 4}}\, \delta_{s_0+s_0',0}^{(2)}, \cr
        {\cal P}_{(l,m,s),(l',m',s')}&={1\over 2\sqrt{2k+4}}\, \sigma_{(l,m,s)}\, 
   \sigma'_{(l',m',s')} \, e^{{i\pi\over 2}{m\, m' \over k+2}}\,
    e^{-i\pi{s\, s' \over 4}}\, \delta_{s+s',0}^{(2)}\cr
    &\left[
    P_{l,l'}\, \delta_{m+m'+k+2,0}^{(2)}+ (-1)^{l'+m'+s'\over 2}\,
     e^{i\pi{m+s\over 2}}\, P_{l,k-l'}\, \delta_{m+m',0}^{(2)}
   \right] .}}
The extra sign factors in \pmatrix,
\eqn\sino{\eqalign{  \sigma_{s_0}&=(-1)^{h_{s_0}-\Delta_{s_0}} \cr
    \sigma_{(l,m,s)}&=(-1)^{h^l_{m,s}-\Delta^l_{m,s}} }}
stem from the roots of the modular $T$-matrix in the definition
of $P$.

Since in the following we restrict ourselves to the case of all
levels being  odd, we present in Table 1 all Gepner models of this type
and their corresponding Calabi-Yau manifold.  
\vskip 0.8cm
\vbox{ \centerline{\vbox{ \hbox{\vbox{\offinterlineskip
\def\tablespace{height2pt&\omit&&\omit&&
 \omit&\cr}
\def\tablerule{\tablespace\noalign{\hrule}\tablespace}

\hrule\halign{&\vrule#&\strut\hskip0.2cm\hfill #\hfill\hskip0.2cm\cr
& levels && $(h_{21},h_{11})$ && CY &\cr
\tablerule
& $(1^9)$ &&  $(84,0)$ && $-$ &\cr
\tablespace
& $(1,1,3,7,43)$ &&  $(67,19)$ &&  $\IP_{1,5,9,15,15}[45]$ &\cr
\tablespace
& $(1,1,3,13,13)$ &&  $(103,7)$ &&  $\IP_{1,1,3,5,5}[15]$ &\cr
\tablespace
& $(1,1,5,5,19)$ &&  $(65,17)$ &&  $\IP_{1,3,3,7,7}[21]$ &\cr
\tablespace
& $(1,1,7,7,7)$ &&  $(112,4)$ &&  $\IP_{1,1,1,3,3}[9]$ &\cr
\tablespace
& $(1,3,3,3,13)$ &&  $(75,3)$ &&  $\IP_{1,3,3,3,5}[15]$ &\cr
\tablespace
& $(3,3,3,3,3)$ &&  $(101,1)$ &&  $\IP_{1,1,1,1,1}[5]$ &\cr
}\hrule}}}} 
\centerline{ \hbox{{\bf
Table 1:}{\it ~~ odd level Gepner models}}} } 
\vskip 0.5cm
\noindent
Apparently, for all levels odd the number of tensor factors
is either five or nine. Therefore  the formulas to be presented in
the following sections are derived under the assumption of $r=5,9$
and all levels $k_j$ odd. However, we would like to point out
that we have evidence for some of them to hold also for the case
of even levels \timo.

\newsec{Modular invariant partition functions from simple current construction}

\subsec{Review of general simple current extension}

Recall that for a given  conformal field theory there 
exists a very general way to construct  modular invariant partition functions
via  an extension of  the chiral symmetry algebra
by some element of the set of simple currents \SchellekensAM.  
These simple currents are primary fields $J_i$ whose OPE
with any other primary field $\phi_i$ only involves one particular other primary 
field, i.e.  
\eqn\sca{ J_a \times \phi_j = \phi_k } 
under fusion. It follows in particular, by associativity of the fusion rules, 
that the OPE of two simple currents yields again a simple current,
so that in a rational CFT the set  of
simple currents forms a finite abelian group ${\cal S}$ under the fusion product. 
Denoting the length of the simple current $J_a$ as  ${\cal N}_a$,
the set $ \{J_a, J_a^2, \ldots, J_a^{{\cal N}_a} \}$ forms 
an abelian subgroup of ${\cal S}$ isomorphic to  ${\ZZ}_{{\cal N}_a}$ with  
$ (J_a^n)^C \equiv (J_a ^{ n}) ^{-1} = J_a ^{{\cal N}_a - n}$. 
Similarly, every simple current groups the primary fields into orbits 
$\{ \phi_i, J_a \times \phi_i, J_a^2 \times \phi_i, 
\ldots, J_a^{{\cal N}^i_a - 1}\times \phi_i\}$ whose length ${\cal
N}^i_a$ is a divisor of ${\cal N}_a$.
 
The crucial observation  is that the action of simple currents in a 
RCFT implies the existence of a conserved quantity for every
primary $\phi_i$, the monodromy charge $Q^{(a)}_i$, defined by
\eqn\scb{ 
J_a(z) \,  \phi_i(w) = (z - w )^{ - Q^{(a)}_i} \phi_k (w)+\dots. }

The monodromy of the identity being 1, it is clear that 
$ Q^{(a)}_i = {t^i_a\over {\cal N}_a}\ {\rm mod}\ 1$ for some integer $t^i_a$.
On the other hand, the monodromy is given by the conformal dimensions of the primary and 
the simple current as
\eqn\scc{ 
Q^{(a)}_i = h (\phi_i) + h ( J_a) - h (J_a \times \phi_i) \ {\rm mod}\ 1, }
from which one can show that 
\eqn\scd{ 
Q ^{(a)}_i(J_a^n \times \phi_i) = {t^i_a + r_a n  \over {\cal N}_a} \ {\rm mod}\ 1.} 
Here the monodromy parameter $r_a$ is defined such that 
\eqn\sce{
h (J_a) = {r_a ({\cal N}_a - 1 ) \over 2{\cal N}_a}  \ {\rm mod}\  1.}

Inspired by the idea of orbifolding the CFT with respect to this world-sheet symmetry, 
one can prove that 
a simple current $J_a$ with even monodromy parameter $r_a$ induces the following even 
more general modular invariant partition function
\eqn\scf{
Z_D ( \tau, \bar{\tau})  = \sum_{k,l} \chi_k (\tau)\,\, (M_a)_{kl}\,\, \chi_l (\bar{\tau}),} 
where 
\eqn\scg{
(M_a)_{kl} = \sum_{p = 1}^{{\cal N}_a} \delta( \phi_k, J_a^p \times \phi_l) \, \, \, 
\delta^{(1)} \big(\hat{Q}^{(a)}(\phi_k) + \hat{Q}^{(a)} (\phi_l) \big) }
and 
\eqn\sch{
\hat{Q}^{(a)} (\phi_i) = {t_a^i \over 2 {\cal N}_a} \ {\rm mod}\ 1 . }
Note that the proof relies on the fact that $r_a$ is even, which can always be arranged 
for odd ${\cal N}_a$, with $r_a$ being defined only ${\rm mod}\, {\cal N}_a$.
This yields two very different types of simple current invariants: 
If the conformal dimension of $J_a$ is integer-valued, i.e. $r_a = 0 \,\,{\rm mod}\,1$, 
$Z_D$ can be written as a left-right symmetric combination of orbits of primaries of 
integer monodromy, each occurring with multiplicity ${{\cal N}_a \over{\cal N}_a^i}$. In all other
cases, $Z_D$ forms an automorphism type invariant in the sense that 
$Z_D = \sum_{l} \chi_l(\tau) \,\,\chi_{\Pi(l)}(\bar{\tau})$ for some permutation $\Pi$
of the primary labels. Furthermore, given two modular invariant matrices 
$M_{a_1}$ and $M_{a_2}$, it is clear that
$Z_D = {1\over N}\sum_{k,l,m,} \chi_l \, (M_{a_1})_{lk} \, (M_{a_2})_{km}\, \chi_m$ 
is another modular invariant partition function with obvious generalizations for 
several $M_{a_i}$; 
the normalization factor $N$ ensures
that the vacuum appears precisely once in $Z_D$.  As can be checked from above, 
$(M_a) ^2 = 1 $. The matrices $M$ are also seen to commute if the respective simple 
currents $J_a$ and $J_b$ are mutually local, i.e. if their 
relative monodromy charge $Q^{(a)} (J_b) = 0  \ {\rm mod}\  1$, in which case 
the various $\delta$-function constraints simplify considerably, as will turn out 
in our application to the Gepner model below.

The above procedure can equally well be used to construct simple current invariants 
for partition functions other than the diagonal invariant.
In particular, one can verify that the extended 
charge conjugated partition function is obtained from 
\eqn\sch{
(M_a^C)_{kl}= \sum_{p=1}^{{\cal N}_a} \, \delta(\phi_k, (J_a^p \times \phi_l)^C)\, 
    \delta^{(1)} \big(\hat{Q}^{(a)}(\phi_k) + 
   \hat{Q}^{(a)}({\phi}_l^C)\big). }

\subsec{Simple currents in the Gepner model}

To begin with, let us identify the simple currents for the Gepner
 model.
Given the fusion rules
\eqn\gepc{
\phi_{ (m_1, s_1)}^0 \times \phi_{ (m_2, s_2)}^{l_2}  =  
          \phi_{ (m_1 + m_2, s_1 + s_2)} ^{l_2} ,}
we conclude that the simple currents $J_{\alpha}$ of the Gepner model under 
consideration can be labeled by the vector
\eqn\gepd{
j_{\alpha} = ( s_0 ^{\alpha}; m_1^{\alpha}, \ldots, m_r^{\alpha}; s_1^{\alpha}, 
    \ldots, s_r^{\alpha}). }

To compute the orbit length of the generic primary $\phi_{ (m_i, s_i)}^{l_i}$ under 
$ J_{\alpha}$, we take into account that the $m_i$- and 
$s_i$- labels are only defined ${\rm mod} \, (2k_i + 4)$ and ${\rm mod} \, 4$ 
respectively, so that for all levels odd (i.e. in the absence of short orbits resulting
from the field identifications 
$(l_i, m_i, s_i) \cong (k_i - l_i, m_i + k_i +2, s_i + 2)$) 
the orbit length ${\cal N}_{\lambda, \mu}^{\alpha} $ is the smallest positive integer 
such that 
\eqn\gepe{\eqalign{
x_i ^{\alpha}  = {{\cal N}_{\lambda,\mu}^{\alpha} \over 2( k_i +2)} \, m_i^{\alpha}  
\quad {\rm and} \quad
 y_j ^{\alpha}  = {{\cal N}_{\lambda,\mu}^{\alpha} \over 4} \, s_j^{\alpha} }}
are integer-valued numbers for all $i, j$ from $\{1, \ldots, r\}$ and $\{0, \ldots, r\}$ 
respectively. 
It is clear that the orbit length only depends on the simple current in question 
and is the same for every primary, ${\cal N}_{\lambda, \mu}^{\alpha}
\equiv {\cal N}^{\alpha} $.

As indicated above, the simultaneous extension of the chiral algebra by several 
simple currents is possible as long as they are mutually local.
For the construction to make sense, we therefore need to impose first compatibility 
with the simple currents \mvec\ generating the GSO-like Gepner projection, 
i.e. 
\eqn\gepf{\eqalign{
J_0  \quad  {\rm with} \quad j_0 \equiv \beta_0 = (1;1, \ldots, 1; 1, \ldots, 1)  \cr
J_i \quad {\rm with}  \quad j_i \equiv \beta_i = ( 2; 0, \ldots, 0; 0,\ldots, 
\underbrace{ 2}_{i-th}, \ldots, 0). }}

Starting with the latter and using \scc\ together with \quannum\ for the computation of the monodromy charge, 
this yields the constraints
\eqn\gepg{
Q^{(\alpha)}( J_i) = - {s_0^{\alpha}\over 2} - {s_i^{\alpha}\over 2} = 
   0 \quad {\rm mod}\ 1,}
so that $s_0^{\alpha} + s_i^{\alpha} = 0 \quad {\rm mod}\ 2$ for $i =
   1,\ldots r$ . 
Without loss of generality, we can restrict ourselves to the case of even
$s_0^{\a}, s_i^{\a}$ and therefore also even $m_i^{\a}$,
 as required by the usual constraint that
$l_i^{\alpha}+m_i^{\alpha}+s_i^{\alpha} = 0 \quad {\rm mod}\ 2$. The point is that
this is not really a restriction of the generality of the currents to
be used since currents from the Neveu-Schwarz sector are  related to those in the Ramond sector by spectral flow, of course.

Furthermore,  we can even assume for the following reason that actually 
$s_i^{\alpha} = 0 \quad {\rm mod}\ 4$. 
Since we will eventually be summing over the orbit generated by $J_i$ and $J_{\alpha}$ 
in the partition function independently (cf.  (3.20) and (3.22) below),
a non-zero value of $s_i^{\alpha}$ can be compensated for by the corresponding contribution 
from $J_i$. In other words, we can encode the information
about the value of $s_j^{\alpha}$ by a shift of $s_0^{\alpha}$, 
because with the above procedure the two currents
$(s_0^{\alpha}; m_j^{\alpha}; 2)$ and $(s_0^{\alpha} +2; m_j^{\alpha}; 0)$ are equivalent.

Mutual locality with the Gepner current $J_0$ necessitates  that
\eqn\geph{
Q^{(\alpha)} (J_0)= \sum_{i=1}^{r} \left( {m_i^{\alpha} \over 2( k_i + 2 )}  \right) - 
 {s_0^{\alpha} \over 4} = 0 \quad {\rm mod} \,1.  }
For all levels $k_i$ odd, this can only be satisfied if $s_0^{\a} = 0
 \quad {\rm mod} \, 4$.
To summarize, the most general set of simple currents consistent with
 the Gepner extension is labeled by the vector
\eqn\gepz{ j_{\a} = (0; m_1^{\a}, \ldots, m_r^{\a}; 0, \ldots,0) }
for even $m_i^{\a}$. In view of \gepe, this means that the orbit
length is odd for every simple current under consideration.

Furthermore, compatibility of the simple currents with each other imposes the constraint
\eqn\gepi{
Q_{\alpha, \beta} \equiv Q^{(\alpha)}(J_{\beta}) = \sum_{i=1}^{r} 
 \left( {m_i^{\alpha} \, m_i^{\beta} \over 2( k_i + 2 )} \right)= 0 
\quad  {\rm mod} \ 1  \quad {\rm for} \quad \alpha \not= \beta. }

Finally, it is sufficient to consider the cases in which none of the simple currents 
lies in the cyclic group generated by any of the other simple
currents or of an arbitrary product of them. As pointed out above, we do not miss any 
generically new cases by this requirement, since, e.g.,
$ M(J_{\alpha})\,  M(J_{\alpha}^n) = 1$ for all $ n \not= 0$.

\subsec{Diagonal and charge conjugate partition functions for Gepner models}

We are now in a position to construct the diagonal invariant for the Gepner model 
extended by appropriate currents $J_{\alpha}, \alpha = 1, \ldots, I$.
Since ${\cal N}_{\alpha}$ is odd, $J_{\alpha}$ and $J_{\alpha}^2$ generate the same orbit 
when acting  on the primary $\phi_{ \mu}^{\lambda}$, which may therefore be
parametrized  as  $\{\phi_{ \mu  +  2\tau_{\alpha} j_{\alpha} }^{\lambda}\}$
with $\tau_{\alpha} = 0, \ldots,{\cal N}_{\alpha}-1$.

To deal with the $\delta$-functions appearing in the extended partition function, 
we observe that the arguments of the $\delta^{(1)}$-functions as
written in  \scg\ are of the form 

\eqn\gepj{\eqalign{
\hat{Q}^{(\alpha)} \left(\phi^{\lambda}_{\tilde{\mu} + 2 \,\tau_{\alpha} j_{\alpha}}\right) + 
 \hat{Q}^{(\alpha)} \left(\phi^{\lambda}_{\tilde{\mu}} \right)  =    
{Q}^{(\alpha)} \left(\phi^{\lambda}_{\tilde{\mu}} \right)+ 
 2\, \tau_{\alpha} \hat{Q}^{(\alpha)} (J_{\alpha})  =
 \left({Q}^{(\alpha)} \left(\phi^{\lambda}_{\tilde{\mu}} \right)+ 
 2\, \tau_{\alpha} {r_{\alpha} \over 2 {\cal N}_{\alpha}} \right) \quad {\rm mod} \ 1, }}
where $\tilde{\mu} = \mu + \sum_{\beta \not= \alpha} 2\, \tau_{\beta} \, j_{\beta}$ 
accounts for the twists from the simple currents other than $\tau_{\alpha}$.
A straightforward calculation shows that due to mutual locality of the currents
\eqn\gepk{
{Q}^{(\alpha)} \left(\phi^{\lambda}_{\mu + \sum_{\beta \not= \alpha}2\,\tau_{\beta}\, 
j_{\beta}}\right) = {Q}^{(\alpha)} \left(\phi^{\lambda}_{\mu} \right) \quad {\rm mod} \ 1. }
Putting things together results in the following
extension of the  diagonal 
invariant partition function \parti\

\eqn\gepm{\eqalign{
Z_D (\tau, \bar{\tau})  &= {1\over N}  {1\over 2^r} 
 {({\rm Im} \tau)^{ - 2}\over |\eta(q)|^4} \sum_{b_0 = 0}^{K-1}
\sum_{b_1, \ldots, b_r = 0 }^{1} 
\sum_{\tau_1= 0}^{{\cal N}_1 -1} \ldots \sum_{\tau_I= 0}^{{\cal N}_I -1}
{\sum_{\lambda,\mu}}^{\beta}\,  
( -1)^{s_0} \cr
&\prod_{\alpha= 1}^I \, \delta^{(1)} \left( Q^{(\alpha)}_{\lambda,\mu}  + 
2\, \tau_{\alpha} \hat{Q}^{\alpha}(J_{\alpha}) \right)
 \, \chi^{\lambda}_{\mu}(q) \, \,   
\chi^{\lambda}_{\mu + b_0 \beta_0 + b_1 \beta_1 + \ldots + 
b_r \beta_r + \sum_{\alpha} 2\,\tau_{\alpha} j_{\alpha}  }(\bar{q})  ,}}
where 
\eqn\gepn{
Q^{(\alpha)}_{\lambda,\mu} =  Q^{(\alpha)} ( \phi^{\lambda}_{\mu} ) = 
\sum_i \left( {m_i\, m_i^{\alpha} \over 2 ( k_i + 2)}\right) \quad {\rm mod} \ 1 . }

Likewise, one obtains a similar expression for the charge conjugate partition function
\eqn\gepo{\eqalign{
Z_C (\tau, \bar{\tau})  &= {1\over N} {1\over 2^r} 
{( {\rm Im} \tau)^{ - 2}\over | \eta (q)|^2 }
 \sum_{b_0 = 0}^{K-1}
\sum_{b_1, \ldots, b_r = 0 }^{1} 
\sum_{\tau_1= 0}^{{\cal N}_1 -1} \ldots \sum_{\tau_I= 0}^{{\cal N}_I -1}
{\sum_{\lambda,\mu}}^{\beta}\,
( -1)^{s_0} \cr
&\prod_{\alpha= 1}^I \, \delta^{(1)} \left( Q^{(\alpha)}_{\lambda, - \mu}  + 
2 \,\tau_{\alpha} \hat{Q}^{(\alpha)}(J_{\alpha}) \right)
\, \chi^{\lambda}_{\mu}(q) \, \,  
\chi^{\lambda}_{-\mu + b_0 \beta_0 + b_1 \beta_1 + \ldots + b_r \beta_r + 
\sum_{\alpha} 2\, \tau_{\alpha} j_{\alpha}  }(\bar{q}) .}}

\newsec{Orientifolds of extended Gepner models: The A-type Klein bottle}

A-type orientifolds are obtained by projecting the bulk theory as
defined by the charge conjugate partition function onto
$\Omega$-invariant states, i.e. onto those states coupling
symmetrically in $Z_C$.
From \gepo\ it is evident that  this projection requires for states 
appearing in the Klein bottle $K^A$
\eqn\kaa{
\mu \cong  -\mu + b_0 \beta_0 + b_1 \beta_1 + \ldots + b_r \beta_r + \sum_{\alpha}2\, 
\tau_{\alpha} j_{\alpha},    }
i.e.
\eqn\kab{
m_j = b + \sum_{\alpha} \tau_{\alpha} m_j ^{\alpha}  \quad {\rm mod} \ (k_j +2) 
\quad {\rm for\ all}\  j }
for some $b$ in the range $\{0, \ldots, \frac{K}{2} -1 \}$ and  where $\tau_{\alpha}$ 
is as usual from $ \{0, \ldots, {\cal N}_{\alpha} -1\}$ but
has to satisfy in addition the $\delta$-constraints from $Z_C$, i.e. is actually a 
function $\tau_{\alpha} ( \lambda, \mu)$. 

To incorporate these projections correctly, it is convenient to read off the 
$\delta$-constraints directly from the very general from
of the simple current extended conjugate partition function \sch\  as
\eqn\kac{\eqalign{
\prod_{{\gamma}= 1}^I \delta^{(1)} \left( \hat{Q}^{(\gamma)}_{-\mu + \sum_{i=0}^r b_i \beta_i + 
\sum_{\alpha=1}^{\gamma -1} 2 \tau_{\alpha}j_{\alpha} } + 
\hat{Q}^{(\gamma)}_{-\mu + \sum_{i=0}^r b_i \beta_i + \sum_{\alpha=1}^{\gamma} 
2\tau_{\alpha} j_{\alpha} } \right). }}

Under the Klein bottle projection  \kaa, a generic $\delta$-constraint is seen to equal
\eqn\kad{\eqalign{
&\delta^{(1)} \bigg( \hat{Q} ^{(\gamma)}_{- \mu + \sum_{i=0}^r b_i \beta_i + 
\sum_{\alpha=1}^{\gamma-1}  2\tau_{\alpha} j_{\alpha} }+ 
 \hat{Q} ^{(\gamma)}_{- \mu + \sum_{i=0}^r b_i \beta_i + \sum_{\alpha=1}^{\gamma} \, 
2 \tau_{\alpha} j_{\alpha} }\bigg) = \cr 
&\delta^{(1)} \bigg( \hat{Q} ^{(\gamma)}_{- \mu + \sum_{i=0}^r b_i \beta_i + 
\sum_{\alpha=1}^{\gamma-1} \, 2 \tau_{\alpha} j_{\alpha} }+ 
 \hat{Q} ^{(\gamma)}_{ \mu - \sum_{\alpha =\gamma +1}^{I} \, 2 \tau_{\alpha}j_{\alpha} }\bigg) \cr
&= \delta^{(1)} \bigg(\sum_{i=0}^r b_i \hat{Q}^{(\gamma)} ( J_i) + 
\sum _{\alpha=1}^{\gamma-1}\tau_{\alpha} Q^{(\gamma)}(J_{\alpha})
- \sum_{\alpha = \gamma +1}^{I}  \tau_{\alpha} Q^{(\gamma)}(J_{\alpha}) \bigg). }}

Two things are crucial to observe: First, the projection on states with integer monodromy 
drops out completely, 
 since, of course, $\hat{Q}^{(\gamma )} _{\mu} + \hat{Q}^{(\gamma )} _{ -\mu} = 0 $. Second, and most 
importantly, the (possibly non-integer) monodromy charges of the currents  
with themselves, $Q^{(\alpha)}(J_{\alpha})$, do not occur in any of the arguments either. 
As for the hatted  monodromies of the Gepner currents, it is clear 
that the monodromy charge of the currents $J_i$, $i=1, \ldots,r$ vanishes anyway
because $s_i^{\alpha}$ vanishes mod 4 for all $i$. The Klein bottle projection \kaa\ implies 
furthermore that only even values of $b_0$ contribute,
so that for $J_0$ we are left with the unhatted monodromy charge in \kad\  
as well, which is integer a priori.

Therefore, the
monodromy projections are satisfied identically and we are left with the following A-type
Klein bottle
\eqn\kae{
K^A = 4 \int_0^{\infty} \frac{dt}{t^3} \frac{1}{2^{r+1}} \frac{1}{\eta( 2it)^2} 
{\sum_{\lambda,\mu}}^{\beta}\, 
 \sum_{b=0}^{\frac{K}{2}-1} \,  
\sum_{\tau_1= 0}^{{\cal N}_1 -1} \ldots \sum_{\tau_I= 0}^{{\cal N}_I -1} \,\,
\prod_{j=1}^r \, \delta^{(k_j +2)}_{m_j, b + \sum_{\alpha} \tau_{\alpha} m_j^{\alpha}} 
(-1)^{s_0} \,
 \chi^{\lambda}_{\mu} ( 2 i t ). }
Transforming this into tree channel with the methods of \BlumenhagenSU\ yields
\eqn\kaf{\eqalign{
\tilde{K}^A &= \frac{ 2^4 \big( \prod_{\alpha=1}^I {\cal N}_{\alpha}\big) }
{ 2^{\frac{3r}{2}} \prod_j \sqrt{k_j + 2} }
\int_0^{\infty} dl \frac{1}{\eta^2(2il)} 
{\sum_{\lambda',\mu'}}^{ev}\, 
\sum_{\nu_0 = 0}^{K-1} \,
\sum_{\nu_1, \ldots, \nu_r=0}^{1} \,
\sum_{\epsilon_1, \ldots, \epsilon_r=0}^{1}\  
(-1)^{\nu_0} \cr 
&\delta^{(4)}_{s'_0 + \nu_0 + 2 \sum \nu_j +2,0 }\,\, 
\delta^{(K')}_{ \sum_j \frac{K'}{2k_j +4} (m'_j + \nu_0 + (1- \epsilon_j)(k_j + 2)), 0}  
\left(\prod_{\alpha} \delta^{(1)}(Q^{(\alpha)}_{\lambda', \mu'})\right) \cr
&\prod_{j=1}^r  \bigg( \frac{P^2_{l'_j, \epsilon_j \, k_j}}{S_{l'_j,0}} \,\,
\delta^{(2)}_{m'_j +\nu_0 + (1- \epsilon_j)(k_j + 2),0 } \,\,
\delta^{(4)}_{s'_j+ \nu_0 +2\nu_j + 2(1-\epsilon_j),0} \bigg)\,\,
\chi^{\lambda'}_{\mu'}(2il), }}
with $K'={\rm lcm}(k_j+2)$.
The crosscap state can be extracted from \kaf\ up to additional sign factors 
\eqn\kag{\eqalign{
\big| C \big>_A &= \frac{1}{\kappa^A_c}
{\sum_{\lambda',\mu'}}^{ev}\, 
\sum_{\nu_0 = 0}^{K-1} \,
\sum_{\nu_1, \ldots, \nu_r=0}^{1} \,
\sum_{\epsilon_1, \ldots, \epsilon_r=0}^{1}\  
\Sigma (\lambda', \mu', \nu_0, \nu_j, \epsilon_j, m_j^{\alpha}) \cr
&\delta^{(4)}_{s'_0 + \nu_0 + 2 \sum \nu_j +2, 0 }\,\, 
\delta^{(K')}_{ \sum_j \frac{K'}{2k_j +4} (m'_j + \nu_0 + (1- \epsilon_j)(k_j + 2)), 0}  
\left(\prod_{\alpha} \delta^{(1)}(Q^{(\alpha)}_{\lambda', \mu'})\right) \cr
&\prod_{j=1}^r  \bigg( \frac{P_{l'_j, \epsilon_j \, k_j}}{\sqrt{S_{l'_j,0}}} \,\,
\delta^{(2)}_{m'_j +\nu_0 + (1- \epsilon_j)(k_j + 2),0 } \,\,
\delta^{(4)}_{s'_j+ \nu_0 +2\nu_j + 2(1-\epsilon_j),0} \bigg)\,\,
\big|{\lambda'},{\mu'}\big>\big>_c,}}
with 
\eqn\kah{
\left(\frac{1}{\kappa_c^A}\right)^2 = 
\frac{ 2^5 \big( \prod_{\alpha=1}^I {\cal N}_{\alpha}\big) }{ 2^{\frac{3r}{2}}K
 \prod_j \sqrt{k_j + 2} } .}

For completeness, we give a similar derivation of the B-type Klein
bottle in the appendix.

\newsec{The A-type Annulus amplitude}

As we can see from the various constraints in  \kaf, it is the states coupling 
diagonally in $Z_D$ which contribute to the divergent A-type
Klein bottle amplitude. For one-loop consistency of the string spectrum we 
therefore  need to introduce an appropriate amount of D-branes, 
i.e. A-type RS boundary states canceling the divergent terms from the orientifold 
plane. The A-type boundary states of the pure Gepner model 
read
\eqn\anal{\eqalign{
 \big|a \big>_A = \big|S_0; (L_j, M_j, S_j)_{j=1}^{r}\big>_A &=
\frac{1}{\tilde{\kappa}_{a}^A} 
{\sum_{\lambda',\mu'}}^{\beta}\,  
(-1)^{\frac{s'^2_0}{2}} e^{ -i\pi \frac{s'_0 S_0}{2}} \cr
&\prod_{j=1}^r \bigg( \frac{S_{l'_j, L_j}}{\sqrt{S_{l'_j,0}}} \,\,
e^{i \pi \frac{m'_j M_j}{k_j +2} } \, e^{-i \pi \frac{s'_j S_j}{2} }\bigg) 
\big|{\lambda',\mu'}\big>\big>, }}
where we use the normalization as computed in \BlumenhagenSU\
\eqn\andl{
\frac{1}{\left(\tilde{\kappa}_{a}^A\right)^2 } = 
\frac {K}{2^{\frac{r}{2} +1} \prod_j \sqrt{k_j +2} }. }
The authors of \FuchsGV\ pointed out that two boundary states of a
theory whose symmetry algebra is extended by a certain group of simple
currents  are equivalent if they lie in the same orbit under a
simple current. Applied to the pure Gepner case, this means that
 the boundary states related by the action of the simple currents
$J_0$ and $J_i$ are equivalent, which is consistent also with the
detailed form of the various open string amplitudes as calculated in
\BlumenhagenSU. Together with the constraint $L_j+M_j+S_j=0$ mod $2$
and the reflection symmetry $(L_j,M_j,S_j)\to (k_j-L_j,M_j+k_j+2,S_j+2)$
this allows us to bring the independent boundary states to the form
$\big|S_0; (L_j, M_j, 0)_{j=1}^{r}\big>_A$ with $L_j=M_j=0$ mod $2$.
Recall from \BlumenhagenSU\ that boundary states with odd $S_j$ are
excluded in the orientifold, being
inconsistent with the M\"obius amplitude, since the crosscap state is
formally given by $\big|0,(L_j,0,0)_{j=1}^r\big>$ for certain numbers
$L_j$ and all tensor factors need to lie
either in the NS- or the R-sector.

In this paper we impose the stronger condition that all boundary states
should also be relatively supersymmetric with respect to the
orientifold plane, thus  giving rise to the additional condition
\eqn\susyg{   \sum_{j=1}^r {M_j\over k_j+2} -   \sum_{j=1}^r {S_j\over 2} - {S_0\over 2}=0
\ {\rm mod}\ 2 .}
With all levels being odd, this implies $S_0=0$ with $S_0=2$ describing the
anti-branes. 
Therefore, the independent supersymmetric A-type boundary states of the pure Gepner model
are given by the boundary states
\eqn\indb{  \big|0; (L_j, M_j, 0)_{j=1}^{r}\big>_A\quad\quad {\rm with\ } L_j,M_j\ {\rm even} .}

For a Gepner model extended by additional simple currents
the boundary states of the new model are given by orbits under the simple
current actions of the boundary states of the pure Gepner model \brunschom\
\eqn\extenboun{   \big|a, J_1,\ldots, J_I\big>_A={1\over \sqrt{
\prod_{\alpha} {\cal N}_\alpha }}\,
                  \sum_{\tau_1=0}^{{\cal N}_1-1}\dots
                    \sum_{\tau_I=0}^{{\cal N}_I-1}
                    \prod_{\alpha=1}^I J_\alpha^{\tau_\alpha}\big|a\big>_A . }
Therefore, now the independent boundary  states are labeled by simple current
orbits of the pure Gepner model boundary states. For simplicity, in the following we will
label them by one of the representatives appearing in  the respective  simple current orbit. 
Inserting the Gepner model boundary states \anal\ into \extenboun, the latter
ones can also be written as
\eqn\ana{\eqalign{
 \big|a\big>_A = \big|S_0; (L_j, M_j, S_j)_{j=1}^{r}\big>_A &=
\frac{1}{\kappa_{a}^A} 
{\sum_{\lambda',\mu'}}^{\beta}\, 
\prod_{\alpha} \delta^{(1)} \big( Q^{(\alpha)}_{\lambda', \mu'} \big) 
(-1)^{\frac{s'^2_0}{2}} e^{ -i\pi \frac{s'_0 S_0}{2}} \cr
&\prod_{j=1}^r \bigg( \frac{S_{l'_j, L_j}}{\sqrt{S_{l'_j,0}}} \,\,
e^{i \pi \frac{m'_j M_j}{k_j +2} } \, e^{-i \pi \frac{s'_j S_j}{2} }\bigg) 
\big|{\lambda',\mu'}\big>\big> }}
with the normalization
\eqn\and{
\frac{1}{\left(\kappa_{a}^A\right)^2 } = 
\frac {K \,\left(\prod_{\alpha} {\cal N}_{\alpha}\right)}{2^{\frac{r}{2} +1} \prod_j \sqrt{k_j +2} }. }
Therefore, in the boundary states only A-type Ishibashi states with 
integer monodromy charge appear. This is consistent with the analogous
statement for the crosscap states, which clearly holds in view of \kag.

Forming the overlap between two stacks of branes $\big|a\big>$ and 
$\big|\tilde{a}\big>$ with Chan-Paton multiplicities $N_{a}$ and
$N_{\tilde{a}}$ respectively results in the tree-channel annulus amplitude
\eqn\anb{\eqalign{
\tilde{A}_{\tilde{a}a}^A &= \frac{N_{a} 
N_{\tilde{a}}}{\kappa_{a}^A\,  \kappa_{\tilde{a}}^A }
\int_0^{\infty}\frac{dl}{\eta^2(2il)} \,
{\sum_{\lambda',\mu'}}^{\beta}\,
\prod_{\alpha} \delta^{(1)} \big( Q^{(\alpha)}_{\lambda', \mu'} \big) 
(-1)^{s'^2_0} 
e^{ -i\pi \frac{s'_0}{2} (S_0-\tilde{S}_0) } 
 \times \cr
&\quad \prod_{j=1}^r \bigg( \frac{S_{l'_j, L_j} \, S_{l'_j,\tilde{L}_j}}{S_{l'_j,0}} \,\,
e^{i \pi \frac{m'_j (M_j- \tilde{M}_j)}{k_j +2} } \, 
e^{-i \pi \frac{s'_j (S_j-\tilde{S_j})}{2} }\bigg)
\chi^{\lambda'}_{\mu'}(2il).  }}
   
In order to finally read off the massless spectrum, we have to transform into loop channel
\eqn\anc{\eqalign{
A_{\tilde{a}\, a}^A &= N_{a} N_{\tilde{a}}\, {1\over 2^{r+1}}\,
\int_0^{\infty} \frac{dt}{t^3} \frac{1}{\eta^2(it)} 
 {\sum_{\lambda,\mu}}^{ev}\,
\sum_{\nu_0 = 0}^{K-1}
\sum_{\nu_1, \ldots, \nu_r = 0 }^{1}
\sum_{\epsilon_1, \ldots,\epsilon_r=0}^1 
\sum_{\sigma_1= 0}^{{\cal N}_1 -1} \ldots \sum_{\sigma_I= 0}^{{\cal N}_I -1}  
( -1)^{\nu_0} \cr
&\delta^{(4)}_{s_0, 2+ \tilde{S}_0 - S_0 - \nu_0 - 2 \sum_j \nu_j} 
\prod_{j=1}^r \bigg( N^{|\epsilon_j k_j - l_j|}_{L_j, \tilde{L}_j}\,\,
\delta ^{(2k_j +4)}_{m_j + M_j - \tilde{M}_j + \nu_0 + \sum_{\alpha} 
\sigma_{\alpha} m_j^{\alpha} + \epsilon_j (k_j+2), 0} \cr
&\delta^{(4)}_{s_j, \tilde{S}_j - S_j - \nu_0 - 2 \nu_j + 2 \epsilon_j} \bigg) 
\chi_{\mu}^{\lambda} (it).  }}

\newsec{The A-type M\"obius amplitude}

In \BlumenhagenSU, a strategy to determine the signs of the crosscap states was developed: 
Examining the transformation of a whole orbit generated by
simple currents (in the case studied there,  the Gepner currents) under world-sheet duality, 
it turned out to be possible to fix the signs of $\big|C\big>$ by requiring that
each orbit transform exactly into another full orbit. 
This yields the M\"obius amplitude as the overlap between crosscap and RS boundary states 
up to an overall sign, which, as is usually done in constructing the open string sector, 
is fixed a posteriori by the tadpole conditions.
Following this general philosophy, it is important to take the distinction between simple 
currents of integer and non-integer conformal dimension
into consideration. Recall that the latter act as automorphisms on the set of primaries 
in the diagonal invariant, as described in section 3, so that the actual
orbits are generated only by the integer spin currents and, of course, the currents inducing 
the Gepner projection as before.
Suppose therefore  that $h(J_\alpha) = 0 \quad {\rm mod}\ 1 $ for all $\alpha$ in 
${ 1, \ldots, I'}$. We are then interested in the P-transformation  of the hypothetical 
Neveu-Schwarz sector M\"obius amplitude
\eqn\moa{
M^{\lambda}_{\mu} = \sum_{\nu_0 = 0}^{ \frac{K}{2} } \, \sum_{ \nu_1, \ldots, \nu_r = 0}^{1} 
\sum_{\tau_1=0}^{{\cal N}_1-1} \ldots \sum_{\tau_{I'}=0}^{{\cal N}_{I'}-1} \, 
(-1)^{[h^{\lambda}_{\mu} (\nu_0, \nu_j, \tau_{\alpha}) - h^{\lambda}_{\mu} ]} \,\,
\hat{\chi}^{\lambda}_{\mu + 2 \nu_0 \beta_0 + \sum \nu_j \beta_j +  
\sum_{\alpha=1}^{I'} 2 \tau_{\alpha} m_j^{\alpha} } (it + \frac{1}{2}),  }
where $h^{\lambda}_{\mu}(\nu_0, \nu_j, \tau_{\a})$ are the conformal
dimensions of the states appearing in the orbit and $h^{\lambda}_{\mu}
= h^{\lambda}_{\mu}(0,0,0)$ and the extra signs go back to writing
the amplitude in terms of the hatted characters. 
The latter ones are defined as
\eqn\realm{  \hat\chi(it+1/2)=e^{-i\pi \left(h-{c\over 24}\right)}\, \chi(it+1/2) .} 
Note that the primary 
$\phi^{\lambda}_{\mu}$ from which the orbit is generated
 has to appear in the partition function, of course, to be eligible for the M\"obius amplitude 
at all, i.e. it has to satisfy the Gepner constraints and must have integer monodromy with the 
integer-spin simple currents.  The only open question is in as far the simple current 
contribution to the total conformal  dimension of the states modifies the resummation of 
the P-transformation.  As is shown in the appendix,
the signs turn out to be  unchanged  as compared 
to \BlumenhagenSU. 
In particular, the orbit-into-orbit condition is seen to be satisfied correctly, including the 
requirement of integer monodromy of the orbits, and we find eventually
\eqn\mob{\eqalign{
\tilde{M}^{\lambda}_{\mu} &\sim 
{\sum_{\lambda',\mu'}}^{\beta}\,
 \sum_{\epsilon_1, \ldots, \epsilon_r=0}^{1} 
\left(\prod_{\alpha=1}^{I'} \delta^{(1)} ( Q^{(\alpha)}_{\lambda', \mu'} ) \right)
\bigg( \prod_{j=1}^r \sigma_{(l'_j, m'_j, s'_j)} \bigg)
\bigg( \prod_{k<l} (-1)^{\eta_k \eta_l} \bigg) \, \, 
\delta^{(2)}_{\sum_j \eta_j, 0} \, \, 
e^{ - i \pi \frac{s_0 s'_0}{4}} \cr
&\delta^{(2)}_{s_0 + s'_0, 0 }  \, \, 
\prod_{j=i}^r \bigg( P_{l_j, |\epsilon_j k_j - l'_j|} \, 
e^{ i \pi \frac{m_j m'_j}{2k_j +4}} \, \, 
\delta^{(2)}_ {m_j + m'_j + ( 1-\epsilon_j)(k_j +2), 0 } \, \,
e^{ -i \pi \frac{s_j s'_j}{4}} \, \, 
\delta^{(2)}_{s_j +s'_j, 0} \cr
&(-1)^{\epsilon_j \frac{m_j+s_j}{2}}   \, \, 
(-1)^{\epsilon_j \frac{l'_j+ m'_j+ s'_j}{2}} \bigg) 
\hat{\chi}^{\lambda'}_{\mu' } (2il + \frac{1}{2}), }}  
where 
\eqn\moc{
\eta_j = \frac{s_0 +s_j}{2} - \frac{s'_0 +s'_j}{2} + \epsilon_j +1. }
Extracting the signs from \mob\  and forming the overlap between a crosscap and RS 
boundary state in the NS-sector yields for the A-type
\eqn\mod{\eqalign{
\tilde{M}_{a}^{ A, NS} &  =
  - \frac{2 N_a}{\kappa_C^A \kappa_{a}^A}  
\int_{0}^{\infty} dl \frac{1}{\hat\eta^2(2il +\frac{1}{2}) }   
{\sum_{\lambda',\mu'}}^{ev}\,
\sum_{\nu_0 = 0}^{\frac{K}{2}-1}
\sum_{\nu_1, \ldots, \nu_r = 0 }^{1} 
\sum_{\epsilon_1, \ldots, \epsilon_r=0}^{1}\cr
&\prod_{\alpha} \delta^{(1)}\left( Q^{(\alpha)}_{\lambda',\mu'}\right)\,
 (-1)^{\nu_0} \, 
\bigg( \prod_{k<l} (-1)^{\nu_k \nu_l} \bigg)
(-1)^{\sum_j \nu_j}\, \, 
e^{ -i \pi \frac{s'_0 S_0}{2}}  \,\,   
\delta^{(4)}_{s'_0 + 2\nu_0 + 2 \sum \nu_j +2, 0 }\cr
&\delta^{(K')}_{ \sum_j \frac{K'}{2k_j +4} (m'_j + 2 \nu_0 + (1- \epsilon_j)(k_j + 2)), 0} 
\prod_{j=1}^r  \bigg( \sigma_{(l'_j, m'_j, s'_j)} \frac{P_{l'_j, \epsilon_j \, k_j} 
S_{l'_j, L_j}}{S_{l'_j,0}} \,\, 
\delta^{(2)}_{m'_j,\,2\nu_0 + (1- \epsilon_j)(k_j + 2) } \,\, \cr
&\delta^{(4)}_{s'_j,\, 2\nu_0 +2\nu_j + 2(1-\epsilon_j)}  
(-1)^{\epsilon_j \frac{m'_j+s'_j}{2}} 
e^{ - i \pi \frac{ m'_j\, M_j }{k_j +2}}\,
e^{i\pi \frac{s'_j S_j}{2} }\bigg)\,\,
\chi^{\lambda'}_{\mu'} ( 2il + \frac{1}{2}). }}
A lengthy calculation gives the following loop-channel M\"obius amplitude
\eqn\moe{\eqalign{
M_{a}^{A, NS} &= (-1)^s N_{a} \, \,
\frac{1}{2^{r+1}}
\int_0^{\infty} \frac{dt}{t^3} \frac{1}{\hat\eta^2(it+\frac{1}{2})} 
{\sum_{\lambda,\mu}}^{ev}\,  \sum_{\nu_0 = 0}^{\frac{K}{2}-1}
\sum_{\epsilon_1, \ldots, \epsilon_r=0}^{1}  
\sum_{\sigma_1 = 0}^{{\cal N}_1 -1} \ldots \sum_{\sigma_{I} =
0}^{{\cal N}_{I} -1}   \cr
&\bigg( \prod_{k<l} (-1)^{\rho_k \rho_l} \bigg)
\delta^{(2)}_{\sum_j \rho_j, 0}\, \, 
\delta^{(2)}_{s_0,0}
\prod_{j=1}^r  \bigg( \sigma_{(l_j, m_j, s_j)} Y^{l_j}_{L_j, \epsilon_j k_j} \,\, 
\delta^{(2k_j +4)}_{2 M_j + m_j   +2 \nu_0+ \sum_{\alpha} \sigma_{\alpha} m_j^{\alpha} - 
\epsilon_j (k_j+2), 0} \cr &\delta^{(2)}_{s_j,0} \,
(-1)^{\frac{\epsilon_j}{2} [2 S_j -s_j -2\epsilon_j]} \,
 (-1)^{\frac{(1- \epsilon_j)}{2} [2M_j -m_j- \epsilon_j (k_j +2)]} \bigg)  \, \,
\hat{\chi}^{\lambda}_{\mu} (it+\frac{1}{2}),     }}
where
\eqn\moe{ r = 4 s +1, }
\eqn\rhoo{  \rho_j={s_0+s_j\over 2} + \epsilon_j-1}
and the $Y$-tensor is defined as
\eqn\ymatr{  Y_{l_1,l_2}^{l_3}=\sum_{l=0}^k  {S_{l_1,l}\, P_{l_2,l} \, P_{l_3,l}\over
                               S_{0,l} } .}

\newsec{Tadpoles and massless spectra}

\subsec{ Tadpole cancellation conditions}

The massless states lead to divergent terms in the Klein bottle, 
the annulus and the M\"obius amplitude.
In light-cone gauge, these are known to  be those states with conformal dimension 
$h = \frac{1}{2}$.
Recall that in the $N=2$ Super-Virasoro minimal model $h$ is bounded from below  by 
half the U(1)-charge, with equality holding
exactly for chiral primaries, so we conclude that the A- type divergent terms stem at most 
from the fields in the (c,c)-ring.
These are exactly the states
\eqn\tada{\eqalign{
(2) \,( 0,0,0)^5\quad    {\rm as\ well\ as}\quad       (0)\, \prod_j (l_j;l_j;0) 
\quad  {\rm with}\quad \sum_j \frac{l_j}{k_j +2} = 1.}}
Besides, the concrete formulas for the various amplitudes put further constraints on the 
chiral fields to actually contribute.

As in the pure Gepner case, we introduce stacks of $N_a$ A-type RS -boundary states 
$\big|0; \prod_j (L_j^a, M_j^a, 0)\big>$ and also
their $\Omega$-image $\big|0; \prod_j (L_j^a, -M_j^a, 0)\big>$. 
One can then check the $\delta$-function constraints in each of the tree-channel
A-Type amplitudes separately, assuming w.l.o.g that $m'_j$ is 
even for the Ishibashi states by 
reflection symmetry 
$(l'_j, l'_j, 0) \cong (k_j - l'_j, l'_j + k_j +2, 2)$.
The result is that only the above chiral fields satisfying in addition 
\eqn\tadb{
Q^{(\alpha)}_{\lambda', \mu'} = 0 \quad {\rm mod}\ 1 }
give a non-vanishing contribution. This is the net effect of the simple current extension and 
reduces the number of tadpole conditions to be satisfied.  
The actual tadpole conditions as such are unaltered as compared to the pure Gepner case 
\BlumenhagenSU\ and 
take the amazingly simple form  
\eqn\tadpolea{\eqalign{  
         \Biggl( \sum_{a=1}^N 2\, N_a\, \cos\biggl[\pi\sum_j 
    {\textstyle{m_j\, M^a_j\over k_j+2}}
               \biggr]\,
          \prod_j \sin (l_j,L^a_j)_{k_j}  
                    -4 
                \prod_j \sin {1\over 2}(l_j,k_j)_{k_j} \Biggr)^2 =0.}}
From the general tadpole cancellation conditions \tadpolea\ it is immediately
clear that there always exists a simple solution to these equations namely
by choosing one stack of D-branes with
\eqn\simplesol{    L_j={k_j\mp1\over 2},\ M_j=0  }
for all $j$ and $k_j=4n_j\pm 1$.
The Chan-Paton factor
is just $N_1=4$ and for $r=5$  leads to a gauge group $SO(4)$
and for $r=9$ to $SP(4)$. 
The interpretation of this solution is that we have just placed 
appropriate D-branes right on top of the
orientifold plane.

\subsec{The gauge sector}

From the loop channel annulus and M\"obius strip amplitudes it is a straightforward
exercise to compute the massless spectrum. Recall that for each boundary state
$|a\rangle=|0;\prod_j (L^a_j,M^a_j,0)\ra$ we have to introduce 
its $\Omega$ image $|a'\rangle=|-0;\prod_j (L^a_j,-M^a_j,0)\ra$.
For each pair of boundary states we have to determine the number
of massless states in the corresponding loop channel amplitudes.

Gauge fields only  arise from open strings stretched between the same
D-branes, as only then does the vacuum state
\eqn\vac{ (2) \,( 0,0,0)^5\ }
appear in $A_{aa}$. 
 If the boundary state $|a\rangle$ is not invariant under
$\Omega$ but mapped to a different state  $|a'\rangle$, this pair
of branes carries a $U(N_a)$ gauge field on its world-volume. 
Consistently, in this case the M\"obius strip amplitude $M_a$ does
not contain the vacuum state. 

If however the boundary state is invariant under $\Omega$, the vacuum state
does arise in the M\"obius strip amplitude $M_a$ and depending on the
sign one obtains  a gauge field of either $SO(2N_a)$ or $SP(2N_a)$.

\subsec{The matter sector}

Additional massless matter can arise from all possible intersections
of the boundary states.
One has  to compute how many massless states of the form
\eqn\massles{ (0)\, \prod_j (l_j;l_j;0)\quad {\rm respectively\quad } 
              (0)\, \prod_j (l_j;-l_j;0) } 
with $(h,q)=({1\over 2},\pm 1)$
do arise in the annulus and M\"obius strip amplitudes.
More concretely, the various open string sectors give rise to
the chiral massless matter spectrum shown in Table 2. 
\vskip 0.8cm
\vbox{ \centerline{\vbox{ \hbox{\vbox{\offinterlineskip
\def\tablespace{height2pt&\omit&&\omit&&\omit&&
 \omit&\cr}
\def\tablerule{\tablespace\noalign{\hrule}\tablespace}

\hrule\halign{&\vrule#&\strut\hskip0.2cm\hfill #\hfill\hskip0.2cm\cr
& sector &&  $(h,q)$   && rep. && index &\cr
\tablerule
& $A_{ab}$ && $({1\over 2},1)$ && $n^+_{ab}\times (\o{\bf N}_a,{\bf N}_b)$ &&  &\cr
\tablespace
&   && $({1\over 2},-1)$ && $n^-_{ab}\times ({\bf N}_a,\o{\bf N}_b)$ && 
 $I_{ab}=n^+_{ab}-n^-_{ab}$ &\cr
\tablerule
& $A_{a'b}$ && $({1\over 2},1)$ && $n^+_{a'b}\times ({\bf N}_a,{\bf N}_b)$ && &\cr
\tablespace
&   && $({1\over 2},-1)$ && $n^-_{a'b}\times (\o{\bf N}_a,\o{\bf N}_b)$ && 
               $I_{a'b}=n^+_{a'b}-n^-_{a'b}$ &\cr
\tablerule
& $A_{a'a}+M_a$ && $({1\over 2},1)$ && $n^+_{a,{\rm S}}\times {\bf S}_a+n^+_{a,{\rm A}}
 \times {\bf A}_a$  
 &&  $I_{a'a}=(n^+_{a,{\rm S}}+n^+_{a,{\rm A}})-(n^-_{a,{\rm S}}+n^-_{a,{\rm A}})$   &\cr
\tablespace
&   && $({1\over 2},-1)$ && $n^-_{a,{\rm S}}\times \o{\bf S}_a+n^-_{a,{\rm A}}\times 
    \o{\bf A}_a$ && 
     $I_{oa}=(n^+_{a,{\rm S}}-n^+_{a,{\rm A}})-(n^-_{a,{\rm S}}-n^-_{a,{\rm A}})$ &\cr
}\hrule}}}} 
\centerline{ \hbox{{\bf
Table 2:}{\it ~~ massless chiral matter spectrum}}} } 
\vskip 0.5cm
\noindent
Here we have defined the net number of generations by certain  ``topological''
indices which 
correspond to the topological intersection number in the
intersecting brane world models and from the world-sheet point
of view to the Witten index in the corresponding open string sector
respectively. 
Note that for the (anti-)symmetric representations the net number of generations
is given by the following  combination of indices
\eqn\nethum{\eqalign{    n^+_{a,{\rm S}}-n^-_{a,{\rm S}}&={1\over 2}(I_{a'a}-I_{oa}) \cr
                         n^+_{a,{\rm A}}-n^-_{a,{\rm A}}&={1\over 2}(I_{a'a}+I_{oa}) .}}
The index $I_{oa}$ can be considered as the intersection number between
the D-brane $|a\rangle$ and the orientifold plane and is determined
entirely by the M\"obius strip amplitude. 
In addition one finds some adjoint non-chiral matter in the
$A_{aa}$ open string sectors.

\newsec{Examples}

In this section we will exploit the formulas derived in the last sections
for the construction of a number of explicit chiral models.

\subsec{A simple current extension of the $(3)^5$ model}

We start with the simplest Gepner model with levels $(3)^5$, corresponding
to the quintic Calabi-Yau manifold. 
Since $(h_{21},h_{11})=(1,101)$, the pure A-type Gepner Model
orientifold gives rise to
102 tadpole cancellation conditions for the 1984 Chan-Paton factors. 

In order to reduce this to a treatable number we consider the extension
of this model by the two simple currents
\eqn\exta{  J_1=(0;2,-2,0,0,0;0,0,0,0,0), \quad\quad J_2=(0;2,2,-4,0,0;0,0,0,0,0) }
yielding the Hodge numbers $(h_{21},h_{11})=(49,5)$ and therefore leaving only
six tadpole cancellation conditions. Note that the two simple currents
in \exta\ are indeed relatively local.

Now we introduce the possible supersymmetric boundary states. It turns out that, after modding
out the two simple currents and the $\Omega$-action, we are left
with 96 boundary states.
These fall into three categories which can be described as follows.
In the first class there are those 32 boundary states which are
invariant under the $\Omega$ projection
\eqn\extb{ |S^a_0;\prod_j (L^a_j,M^a_j,S^a_j)\ra=|0;\prod_j (L^a_j,0,0)\ra, }
where the $L_j^a$ denote any of the labels listed
in Appendix C. These states carry Chan-Paton indices $N_{3i}$ with $i\in\{0,\ldots,31\}$. The second   class contains the 32 states
\eqn\extc{ |S^a_0;\prod_j (L^a_j,M^a_j,S^a_j)\ra=|0;(L^a_1,-2,0)(L^a_2,0,0)(L^a_3,0,0)
                                              (L^a_4,2,0)(L^a_5,0,0)\ra }
with Chan-Paton indices $N_{3i+1}$.
Finally, the third class are the states
\eqn\extd{ |S^a_0;\prod_j (L^a_j,M^a_j,S^a_j)\ra=|0;(L^a_1,-4,0)(L^a_2,0,0)(L^a_3,0,0)
                                              (L^a_4,4,0)(L^a_5,0,0)\ra }
with Chan-Paton indices $N_{3i+2}$.

The next step is to evaluate  the six tadpole cancellation conditions
in terms of the 96 Chan-Paton factors. After a little algebra one can
bring these conditions into a form with explicitly integer valued
coefficients. 

\noindent
$\bullet$ Condition 1:
\eqn\tada{\eqalign{  12=& 12 N_0 + 2 N_1 + 2 N_2 + 6 N_3 - 4 N_4 + 6 N_5 + 6 N_6 - 4 N_7
+ 6 N_8 + 18 N_9 - 2 N_{10} + 8 N_{11} + \cr
 &6 N_{12} - 4 N_{13} + 6 N_{14} + 18 N_{15} - 
   2 N_{16} + 8 N_{17} + 18 N_{18} - 2 N_{19} + 8 N_{20} + 24 N_{21} - 6 N_{22} + \cr
& 14 N_{23} + 
   6 N_{24} + N_{25} + N_{26} + 8 N_{27} + 3 N_{28} + 8 N_{29}  + 8 N_{30} + 3 N_{31} + 
   8 N_{32} + 14 N_{33} + \cr
&4 N_{34} + 9 N_{35} + 8 N_{36} + 3 N_{37} + 8 N_{38} + 14 N_{39} + 
   4 N_{40}+ 9 N_{41}+ 14 N_{42}+ 4 N_{43}+ 9 N_{44} + \cr
& 22 N_{45}+ 7 N_{46}+ 17 N_{47}+ 
   6 N_{48}+ N_{49}+ N_{50}+ 8 N_{51}+ 3 N_{52}+ 8 N_{53}+ 8 N_{54}+ 3 N_{55}+\cr 
& 8 N_{56}+ 14 N_{57}+ 4 N_{58}+ 9 N_{59}+ 8 N_{60}+ 3 N_{61}+ 8 N_{62}+ 14 N_{63}+ 
   4 N_{64}+ 9 N_{65}+ 14 N_{66}+ \cr
& 4 N_{67}+ 9 N_{68}+ 22 N_{69 }+ 7 N_{70}+ 17 N_{71} - 
   2 N_{72}+ 8 N_{73}+ 8 N_{74}+ 4 N_{75}+ 14 N_{76}+ 4 N_{77}+ \cr
& 4 N_{78}+ 14 N_{79}+ 
   4 N_{80}+ 2 N_{81}+ 22 N_{82}+ 12 N_{83}+ 4 N_{84}+ 14 N_{85}+ 4 N_{86}+ 2 N_{87}+ 
   22 N_{88}+ \cr
&12 N_{89 }+ 2 N_{90}+ 22 N_{91}+ 12 N_{92}+ 6 N_{93}+ 36 N_{94}+ 
   16 N_{95}}}  
\noindent
$\bullet$ Condition 2:
\eqn\tadb{\eqalign{
0=& 5 N_0 + 5 N_1 + 5 N_2 - 3 N_3 - 3 N_4 - 3 N_5 - 3 N_6 - 3 N_7 - 3 N_8 + 2 N_9
    + 2 N_{10} + 2 N_{11} - 3 N_{12} - \cr
& 3 N_{13} - 3 N_{14} + 2 N_{15} + 
 2 N_{16} + 2 N_{17} + 2 N_{18} + 2 N_{19 } + 2 N_{20} - N_{21} - N_{22} - N_{23} - 3 N_{24} -\cr 
&   3 N_{25} - 3 N_{26} + 2 N_{27} + 2 N_{28} + 2 N_{29 } + 2 N_{30} + 2 N_{31} + 2 N_{32} - 
   N_{33} - N_{34} - N_{35} + 2 N_{36} + \cr
&2 N_{37} + 2 N_{38} - N_{39 } - N_{40} - N_{41} - N_{42} - 
   N_{43} - N_{44} + N_{45} + N_{46} + N_{47} - 3 N_{48} - 3 N_{49 } - \cr
& 3 N_{50} + 2 N_{51} + 
   2 N_{52} + 2 N_{53} + 2 N_{54} + 2 N_{55} + 2 N_{56} - N_{57} - N_{58} - N_{59 } + 2 N_{60} + 
   2 N_{61} + \cr
& 2 N_{62} - N_{63} - N_{64} - N_{65} - N_{66} - N_{67} - N_{68} + N_{69 } + N_{70} + 
   N_{71} + 2 N_{72} + 2 N_{73} + 2 N_{74} - \cr
& N_{75} - N_{76} - N_{77} - N_{78} - N_{79 } 
   - N_{80} + 
   N_{81} + N_{82} + N_{83} - N_{84} - N_{85} - N_{86} + N_{87} + N_{88} + \cr
& N_{89} + 2 N_{9} +  N_{90} + N_{91} + N_{92} }}
\noindent
$\bullet$ Condition 3:
\eqn\tadc{\eqalign{
0=&N_{25} - N_{26} - 2 N_{27} + N_{28} - 2 N_{30} + N_{31} - 2 N_{33} + 2 N_{34} - N_{35} - 
   2 N_{36} + N_{37} - 2 N_{39} + 2 N_{40} - \cr
& N_{41} - 2 N_{42} + 2 N_{43} - N_{44} - 4 N_{45} + 3 N_{46} - 
   N_{47} - N_{49} + N_{50} + 2 N_{51} - N_{52} + 2 N_{54} - N_{55} + \cr
& 2 N_{57} - 2 N_{58} + 
   N_{59} + 2 N_{60} - N_{61} + 2 N_{63} - 2 N_{64} + N_{65} + 2 N_{66} - 2 N_{67} + N_{68} + 
   4 N_{69} - \cr
&3 N_{70} + N_{71} }}
\noindent
$\bullet$ Condition 4:
\eqn\tadd{\eqalign{
0=& 2 N_{24} + N_{25} - 2 N_{26} - 2 N_{27} + N_{29} - 2 N_{30} + N_{32} + N_{34} - N_{35} - 
   2 N_{36} + N_{38} + N_{40} - N_{41} + \cr
& N_{43} - N_{44} - 2 N_{45} + N_{46} - 2 N_{48} - 
   N_{49} + 2 N_{50} + 2 N_{51} - N_{53} + 2 N_{54} - N_{56} - N_{58} + N_{59} + \cr
& 2 N_{60} - N_{62} - N_{64} + N_{65} - N_{67} + N_{68} + 2 N_{69} - N_{70} }}
\noindent
$\bullet$ Condition 5:
\eqn\tade{\eqalign{
0=& N_0 + 2 N_{1} + 3 N_{2} - N_{3} - N_{4} - 2 N_{5} - N_{6} - N_{7} - 2 N_{8} + N_{10} + 
   N_{11} - N_{12} - N_{13} - 2 N_{14} + \cr
& N_{16} + N_{17} + N_{19 } + N_{20} - N_{21} - 
   N_{23} - N_{24} - N_{25} - 2 N_{26} + N_{28} + N_{29} + N_{31} + N_{32} - \cr
& N_{33} - 
   N_{35} + N_{37} + N_{38} - N_{39 } - N_{41} - N_{42} - N_{44} - N_{45} + 
   N_{46} - 3 N_{48} - 2 N_{49} + 2 N_{51} + \cr
& N_{52} + 2 N_{54} + N_{55} - N_{57} - N_{58} + 
   2 N_{60} + N_{61} - N_{63} - N_{64} - N_{66} - N_{67} + N_{69} + 2 N_{72} + \cr
& N_{73} - N_{75} - 
   N_{76} - N_{78} - N_{79} + N_{81} - N_{84} - N_{85} + N_{87} + N_{90} - N_{94} }}
\noindent
$\bullet$ Condition 6:
\eqn\tadf{\eqalign{
0=& -N_{1} + 2 N_{3} + N_{5} + 2 N_{6} + N_{8} +2 N_{9} - N_{10} + N_{11} + 2 N_{12} + 
    N_{14} + 2 N_{15} - N_{16} + N_{17} + \cr
& 2 N_{18} - N_{19} + N_{20} + 
   4 N_{21} - N_{22} + 2 N_{23} + 2 N_{24} - N_{25} + 2 N_{26} + 4 N_{27} - 2 N_{28} + N_{29} + \cr
 & 4 N_{30} - 2 N_{31} + N_{32} + 6 N_{33} - 3 N_{34} + 3 N_{35} + 4 N_{36} - 2 N_{37} + N_{38} + 
   6 N_{39} - 3 N_{40} + 3 N_{41} + \cr
& 6 N_{42} - 3 N_{43} + 3 N_{44} + 10 N_{45} - 5 N_{46} + 
   4 N_{47} + 2 N_{49} - 2 N_{51} + 2 N_{52} - 2 N_{54} + 2 N_{55} - \cr
& 2 N_{57} + 4 N_{58} - 
   2 N_{60} + 2 N_{61} - 2 N_{63} + 4 N_{64} - 2 N_{66} + 4 N_{67} - 4 N_{69} + 6 N_{70} + 
   N_{73} + \cr
& 2 N_{76} + N_{77} + 2 N_{79} + N_{80} + 3 N_{82} + N_{83} + 2 N_{85} + N_{86} + 
   3 N_{88} + N_{89} + 3 N_{91} + N_{92} + \cr
&5 N_{94} + 2 N_{95} }}
Clearly, it is not so easy to classify all possible solutions to these six equations. 

Before we display at least a couple of non-trivial solutions, we would like to point
out that the intersection numbers between pairs of the 96 boundary states do not always vanish.
Therefore, in contrast to the B-type orientifold studied in \BlumenhagenSU,  here chiral
models are indeed possible. Moreover, since the intersection number does
not always vanish, we can perform a highly non-trivial test of the
entire presented formalism including the general sign factors
in the M\"obius strip amplitude. 
One can quite generally compute the non-abelian gauge anomaly on a stack
of D-branes of type $N_a$ in terms of all Chan-Paton indices. 
This is given by the following expression
\eqn\gaugeanom{ \sum_{b\ne a} N_b\, \left( I_{a'b}- I_{ab} \right) + 
             (N_a-4){\textstyle{1\over 2}}\left(
               I_{a'a}+I_{oa}\right) + (N_a+4){\textstyle{1\over 2}}\left(
               I_{a'a}-I_{oa}\right) .}
Note that in the definition of the index $I_{oa}$ the signs in the M\"obius strip amplitude 
play a crucial role. Evaluating the 96 gauge anomalies \gaugeanom\ 
and inserting the 6 tadpole cancellation conditions, one can shown that they all
indeed vanish. This constitutes a highly non-trivial test showing that 
everything is correct.

\subsec{A chiral model}

The choice $N_{10}=4$, $N_{17}=2$, $N_{25}=2$ and $N_{49}=2$ with the remaining
Chan-Paton indices vanishing satisfies all six tadpole
cancellation conditions. All four boundary states are not invariant under the
$\Omega$ projection, so that we get a gauge group
\eqn\gaugea{  G=U(4)\times U(2)\times U(2)\times U(2) }
of rank rk$(G)=10$. 
Computing the massless spectrum we find the chiral spectrum displayed
in Table 3.

\vskip 0.8cm
\vbox{ \centerline{\vbox{ \hbox{\vbox{\offinterlineskip
\def\tablespace{height2pt&\omit&&
 \omit&\cr}
\def\tablerule{\tablespace\noalign{\hrule}\tablespace}

\hrule\halign{&\vrule#&\strut\hskip0.2cm\hfill #\hfill\hskip0.2cm\cr
& deg. &&  $U(4)\times U(2)\times U(2)\times U(2)$  &\cr
\tablerule
& $2$ &&  $(1,\o{\bf 2}, {\bf 2},1)$  &\cr
\tablespace
& $1$ &&  $(1,\o{\bf 2}, \o{\bf 2},1)$  &\cr
\tablerule
& $2$ &&  $(1,{\bf 2}, 1, \o{\bf 2})$  &\cr
\tablespace
& $1$ &&  $(1,{\bf 2},1,  {\bf 2})$  &\cr
\tablerule
& $1$ &&  $({\bf 4}, 1, {\bf 2},1)$  &\cr
\tablespace
& $1$ &&  $(\o{\bf 4}, 1, 1, \o{\bf 2})$  &\cr
\tablerule
& $1$ &&  $(1, 1, \o{\bf S},1)$  &\cr
\tablespace
& $1$ &&  $(1, 1, 1, {\bf S})$  &\cr
}\hrule}}}} 
\centerline{ \hbox{{\bf
Table 3:}{\it ~~ massless chiral matter spectrum}}} } 
\vskip 0.5cm
\noindent
Apparently, the non-abelian gauge anomaly cancels. 
This chiral massless spectrum is extended by the non-chiral one in Table 4.
\vskip 0.8cm
\vbox{ \centerline{\vbox{ \hbox{\vbox{\offinterlineskip
\def\tablespace{height2pt&\omit&&
 \omit&\cr}
\def\tablerule{\tablespace\noalign{\hrule}\tablespace}

\hrule\halign{&\vrule#&\strut\hskip0.2cm\hfill #\hfill\hskip0.2cm\cr
& deg. &&  $U(4)\times U(2)\times U(2)\times U(2)$  &\cr
\tablerule
& $1$ &&  $({\bf 4}, \o{\bf 2},1,1)+c.c.$  &\cr
\tablespace
& $3$ &&  $(1,1, {\bf 2}, \o{\bf 2})+c.c.$  &\cr
\tablerule
& $2$ &&  $(1,{\bf S}, 1, 1)+c.c$  &\cr
\tablespace
& $1$ &&  $(1,{\bf A},1, 1)+c.c.$  &\cr
\tablerule
& $5$ &&  $({\bf Adj},1, 1,1)$  &\cr
\tablespace
& $5$ &&  $(1,{\bf Adj},1,1)$  &\cr
}\hrule}}}} 
\centerline{ \hbox{{\bf
Table 4:}{\it ~~ massless non-chiral matter spectrum}}} } 
\vskip 0.5cm
\noindent
Note that the third and fourth stack of branes are rigid in the sense
that there is no additional adjoint matter. 
This is one of the features which are very difficult to realize
in intersecting brane worlds on toroidal orbifolds and which, of course,
are really welcome in the string theoretical realization of the
Standard Model as additional adjoint matter easily spoils asymptotic
freedom  and the nice gauge coupling unification properties 
of the low energy gauge groups \BlumenhagenJY.

If we extend the model further by the  additional simple current
\eqn\bla{J_3=(0;2,2,2,-6,0;0,0,0,0,0)}
 we get the Greene-Plesser mirror
of the quintic and this A-type model simply becomes the 
B-type model studied in \refs{\BlumenhagenTJ,\AldazabalUB,\BlumenhagenSU}.

\subsec{A simple current extension of the $(1)^2\, (7)^3$ model}

As a second non-trivial example we present a model derived
from the $(1)^2\, (7)^3$ Gepner model. The Gepner model
itself has Hodge numbers $(h_{21},h_{11})=(4,112)$ but leads
after extending it by the two simple currents
\eqn\exta{  J_1=(0;2,0,2,0,-8;0,0,0,0,0), \quad\quad J_2=(0;-2,0,6,0,0;0,0,0,0,0) }
to a model with Hodge numbers $(h_{21},h_{11})=(55,7)$.

The possible supersymmetric boundary states come in two classes, where the first
class contains 64 $\Omega$-invariant states and the 64 states in the 
second class can be described as
\eqn\extf{ |S^a_0;\prod_j (L^a_j,M^a_j,S^a_j)\ra=|0;(L^a_1,-2,0)(L^a_2,0,0)(L^a_3,0,0)
                                              (L^a_4,0,0)(L^a_5,0,0)\ra }
with the labels $L^a_j$ taken from the list in Appendix C.
Since we are more interested in unitary gauge groups in the following, 
we only consider these latter boundary states. 
Again after some algebra, the six independent tadpole cancellation conditions
can be brought into a form with only integer coefficients and read:

\noindent
$\bullet$ Condition 1:
\eqn\tada{\eqalign{  20 =&2 N_{0} + 2 N_{1} + 2 N_{2} + 2 N_{4} + 3 N_{5} + 4 N_{6} + 2 N_{7} + 
   2 N_{8} + N_{9} + 3 N_{10} + N_{11} - N_{13} + \cr
   &N_{14} + N_{15} + 2 N_{16} + 6 N_{17} + 4 N_{18} + 
   2 N_{19} + 3 N_{20} + 9 N_{21} + 9 N_{22} + 6 N_{23} + N_{24} + \cr
   &6 N_{25} + 5 N_{26} + 
   4 N_{27} - N_{28} + N_{30} + 2 N_{31} + 2 N_{32} + 4 N_{33} + 6 N_{34} + 4 N_{35} + 4 N_{36} +\cr 
   & 9 N_{37} + 11 N_{38} + 7 N_{39} + 3 N_{40} + 5 N_{41} + 7 N_{42} + 4 N_{43} + N_{44} + 
   N_{45} + N_{46} + 2 N_{49} + \cr
 &4 N_{50} + 4 N_{51} + 2 N_{52} + 6 N_{53} + 7 N_{54} + 
   5 N_{55} + N_{56} + 4 N_{57} + 4 N_{58} + 3 N_{59} + N_{60} + \cr
   &2 N_{61} - N_{63}  \cr
   }}  
\noindent
$\bullet$ Condition 2:
\eqn\tada{\eqalign{ 0 = &2 N_{0} - N_{1} + N_{2} - N_{3} - N_{4} + N_{7} + 2 N_{8} - N_{9} + 
                   N_{10} - N_{11} - N_{16} + 2 N_{17} - N_{18} - \cr
   &N_{21} + N_{22} + 
                   N_{23} - N_{24} + 2 N_{25} - N_{26} + N_{32} - N_{33} + N_{34} + N_{37} + 
                   N_{40} - N_{41} + \cr
    &N_{42} - N_{48} + N_{51} + N_{52} + N_{53} - N_{55} - 
                   N_{56} + N_{59} \cr
   }}  
\noindent
$\bullet$ Condition 3:
\eqn\tada{\eqalign{ -2 = &- N_{0} + N_{1} - N_{2} - N_{5} - N_{8} + N_{9} - N_{10} + 
                   N_{16} - N_{17} - N_{19} - N_{20} - N_{21} - N_{22} + \cr
   &N_{24} - N_{25} - 
            N_{27} - N_{32} - N_{34} - N_{37} - N_{38} - N_{39} - N_{40} - N_{42} - 
            N_{49} - N_{54} - \cr
    &N_{55} - N_{57} \cr
   }}  
\noindent
$\bullet$ Condition 4:
\eqn\tada{\eqalign{ 4 = &- 2 N_{0} + 2 N_{1} - N_{2} + N_{3} + N_{4} + N_{6} - 2 N_{8} + 
                  2 N_{9} - N_{10} + N_{11} + 2 N_{16} - N_{17} + 2 N_{18} +\cr 
   &2 N_{21} + N_{22} + N_{23} + 2 N_{24} - N_{25} + 2 N_{26} - N_{32} + 2 N_{33} + 
   N_{35} + N_{36} +  N_{37} + 2 N_{38} + \cr
 &N_{39} - N_{40} + 2 N_{41} + N_{43} + N_{48} + 
   N_{50} + N_{53} + N_{54} + N_{55} + N_{56} + N_{58} \cr
   }}  
\noindent
$\bullet$ Condition 5:
\eqn\tada{\eqalign{ -4 = &- N_{0} + N_{1} - N_{2} + 2 N_{4} - 2 N_{5} + N_{6} - N_{7} + 
    N_{12} - N_{13} + N_{14} + N_{16} - N_{17} - N_{19} - \cr
  &2 N_{20} + N_{21} - 
    2 N_{22} - N_{28} + N_{29} + N_{31} - N_{32} - N_{34} + N_{36} - 2 N_{37} - N_{39} + N_{44} + \cr
    &N_{46} - N_{49} - N_{52} - N_{54} + N_{61} \cr
   }}  
\noindent
$\bullet$ Condition 6:
\eqn\tada{\eqalign{ -2 = &- 2 N_{0} + 2 N_{1} - N_{2} + N_{3} + 4 N_{4} - 3 N_{5} + 
       3 N_{6} - N_{7} + 
     2 N_{12} - 2 N_{13} + N_{14} - N_{15} + \cr
   &2 N_{16} - N_{17} + 2 N_{18} - 
   3 N_{20} + 4 N_{21} - N_{22} + 2 N_{23} - 2 N_{28} + N_{29} - 2 N_{30} - N_{32} + \cr
   &2 N_{33} + 
   N_{35} + 3 N_{36} - N_{37} + 3 N_{38} + N_{44} - 2 N_{45} - N_{47} + N_{48} + N_{50} - 
   N_{52} + \cr
   &2 N_{53} + N_{55} - N_{60} - N_{62} \cr
   }}  
\noindent
Again one can check quite in general that as long as these
six tadpole cancellation conditions are satisfied the non-abelian
gauge anomalies do all cancel. 

\subsec{A chiral model}

Choosing  $N_{35}=4$, $N_{13}=2$, $N_{15}=2$ and $N_{19}=2$ with the remaining
Chan-Paton indices vanishing satisfies all six tadpole
cancellation conditions. 
The gauge group is the same as in the first example
\eqn\gaugeb{  G=U(4)\times U(2)\times U(2)\times U(2), }
but as shown in Table 5 the chiral massless 
spectrum is completely different.

\vskip 0.8cm
\vbox{ \centerline{\vbox{ \hbox{\vbox{\offinterlineskip
\def\tablespace{height2pt&\omit&&
 \omit&\cr}
\def\tablerule{\tablespace\noalign{\hrule}\tablespace}

\hrule\halign{&\vrule#&\strut\hskip0.2cm\hfill #\hfill\hskip0.2cm\cr
& deg. &&  $U(4)\times U(2)\times U(2)\times U(2)$  &\cr
\tablerule
& $2$ &&  $(\o{\bf 4}, \o{\bf 2},1,1)$  &\cr
\tablerule
& $2$ &&  $(\o{\bf 4}, 1, \o{\bf 2},1)$  &\cr
\tablerule
& $4$ &&  $({\bf 4}, 1, 1, {\bf 2})$  &\cr
\tablerule
& $2$ &&  $(1,{\bf 2},{\bf 2},1)$  &\cr
\tablerule
& $3$ &&  $(1,\o{\bf 2}, 1, \o{\bf 2})$  &\cr
\tablerule
& $1$ &&  $(1,1,\o{\bf 2}, \o{\bf 2})$  &\cr
\tablerule
& $3$ &&  $({\bf A}, 1, 1,1)$  &\cr
\tablespace
& $2$ &&  $(1, {\bf S},1,1)$  &\cr
\tablespace
& $1$ &&  $(1, {\bf A},1,1)$  &\cr
\tablespace
& $1$ &&  $(1, 1, {\bf S},1)$  &\cr
\tablespace
& $1$ &&  $(1, 1, 1, \o{\bf S})$  &\cr
\tablespace
& $1$ &&  $(1, 1, 1, {\bf A})$  &\cr
}\hrule}}}} 
\centerline{ \hbox{{\bf
Table 5:}{\it ~~ massless chiral matter spectrum}}} } 
\vskip 0.5cm
\noindent
As they  should the non-abelian gauge anomalies cancel. 
This chiral massless spectrum is extended by the non-chiral one in Table 6.

These two examples show that A-type orientifolds of Gepner models can lead
to the very characteristics of the (supersymmetric) Standard model like
unitary gauge groups of large enough rank, chirality, three generations
and additional non-chiral (Higgs like) matter. The aim of this paper
was solely to provide the necessary material for dealing with such models
on the technical level. It is clear that a further systematic search has
to be performed in order to really find models which come
closer to the MSSM. 
\vskip 0.8cm
\vbox{ \centerline{\vbox{ \hbox{\vbox{\offinterlineskip
\def\tablespace{height2pt&\omit&&
 \omit&\cr}
\def\tablerule{\tablespace\noalign{\hrule}\tablespace}

\hrule\halign{&\vrule#&\strut\hskip0.2cm\hfill #\hfill\hskip0.2cm\cr
& deg. &&  $U(4)\times U(2)\times U(2)\times U(2)$  &\cr
\tablerule
& $1$ &&  $({\bf 4}, \o{\bf 2},1,1)+c.c.$  &\cr
\tablerule
& $2$ &&  $({\bf 4},1, \o{\bf 2}, 1)+c.c.$  &\cr
\tablerule
& $5$ &&  $({\bf 4},1, 1, \o{\bf 2})+c.c.$  &\cr
\tablespace
& $1$ &&  $({\bf 4},1, 1, {\bf 2})+c.c.$  &\cr
\tablerule
& $1$ &&  $(1,{\bf 2}, \o{\bf 2},1)+c.c.$  &\cr
\tablespace
& $3$ &&  $(1,{\bf 2},  {\bf 2},1)+c.c.$  &\cr
\tablerule
& $2$ &&  $(1,{\bf 2}, 1, \o{\bf 2})+c.c.$  &\cr
\tablerule
& $2$ &&  $(1,1,{\bf 2} , \o{\bf 2})+c.c.$  &\cr
\tablerule
& $2$ &&  $({\bf S}, 1, 1, 1)+c.c$  &\cr
\tablespace
& $2$ &&  $(1,{\bf S},1, 1)+c.c.$  &\cr
\tablespace
& $1$ &&  $(1,1, {\bf S}, 1)+c.c.$  &\cr
\tablespace
& $1$ &&  $(1,1, 1, {\bf S})+c.c.$  &\cr
\tablerule
& $6$ &&  $({\bf Adj},1, 1,1)$  &\cr
\tablespace
& $1$ &&  $(1,{\bf Adj},1,1)$  &\cr
\tablespace
& $1$ &&  $(1,1, {\bf Adj},1)$  &\cr
\tablespace
& $4$ &&  $(1,1,1,{\bf Adj})$  &\cr
}\hrule}}}} 
\centerline{ \hbox{{\bf
Table 6:}{\it ~~ massless non-chiral matter spectrum}}} } 

\newsec{Conclusions}

In this paper we have investigated A-type orientifolds of Gepner models
for their ability to give rise to some of the salient features of the
supersymmetric Standard-Model like unitary gauge groups, chirality, family replication
and large enough gauge groups. After having derived explicitly the general
expressions for all relevant one-loop amplitudes, we have demonstrated by working
out two examples in detail that all the rough Standard-Model features  can be achieved
by simple current extensions of Gepner models. 
This result is very promising and it would be very interesting to
scan the whole plethora of such models for MSSM-like models. 
Given a concrete model, of course one would be interested in 
more refined data like Yukawa couplings or other pieces of the
N=1 low energy effective action like the K\"ahler potential, gauge couplings
and their one-loop threshold corrections.

\vskip 1cm
\centerline{{\bf Acknowledgements}}\pano
The work  of R.B. and T.W. is supported by PPARC and DAAD respectively. 
We  would like to thank Y. Stanev
for helpful  correspondence. 

\vfill\eject
\appendix{A}{B-type Klein bottle projections}

The B-type Klein bottle is computed  from the diagonal partition
function \gepm.
Only those states coupling to one another in $Z_D$ survive the Klein bottle 
$\Omega$-projection which satisfy

\eqn\kba{
\mu \cong \mu + b_0 \beta_0 + b_1 \beta_1 + \ldots + b_r \beta_r + 
\sum_{\alpha}2\, \tau_{\alpha} j_{\alpha},    }
where in addition the $\tau_{\alpha}$ are constrained by the various $\delta$-functions in \gepm.
More abstractly, this is tantamount to requiring that the primaries be fixed points of the 
simple current algebra, i.e.
\eqn\kbb{
\phi^{\lambda}_{\mu} = \prod_{i=0}^{r} (J_i)^{b_i} \, 
\prod_{\alpha=1}^{I} J_{\alpha}^{2\, \tau_{\alpha}}\, \,  
\phi^{\lambda}_{\mu} . }
To put it differently, the product of simple currents on the right-hand side of \kbb\ acts 
as a stabilizer of $\phi^{\lambda}_{\mu}$. 
The set of non-trivial 
stabilizers in a RCFT is empty in the absence of short orbits,
which clearly applies to our discussion of odd integer levels $k_i$. The only possibility 
is therefore that
$\prod_{i=0}^{r} (J_i)^{b_i} \, \prod_{\alpha=1}^{I} J_{\alpha}^{2\, \tau_{\alpha}} = 1 $. 
The case of non-vanishing exponents $b_i, \tau_{\alpha}$
is easily seen to be excluded by the requirement stated above that none of the currents be 
generated by any combination of the others. 
We conclude that the fields appearing in the Klein bottle amplitude are precisely
those for which the $\delta$-functions in $Z_D$ give non-vanishing contributions for 
$\tau_{\alpha} = 0$, 
i.e. those with integer monodromy $ Q^{(\alpha)}_{\lambda,\mu}$. 
This additional projection of the possible states in the Klein bottle is the actual
net effect of the simple current construction.
Thus we find the simple-current extended B-type Klein bottle
\eqn\kbd{
K^B = 4 \int_0^{\infty} {dt\over t^3} \frac{1}{2^{r+1}} \frac{1}{\eta( 2it)^2} 
{\sum_{\lambda,\mu}}^{\beta}\,  \prod_{\alpha=1}^I \bigg(\delta^{(1)} 
( Q^{(\alpha)}_{\lambda,\mu})\bigg)\, \, \chi^{\lambda}_{\mu} ( 2 i t ). }
The loop-channel amplitude is easily transformed into tree-channel by implementing the 
various $\delta$- functions in terms of Lagrange multipliers, e.g. 
\eqn\kbe{
\prod_{\alpha=1}^I  \delta^{(1)} ( Q^{(\alpha)}_{\lambda,\mu}) = 
\left(\prod_{\alpha=1}^I \frac{1}{{\cal N}_{\alpha}}\right) 
\sum_{\sigma_1 = 0}^{{\cal N}_1 -1}\ldots \sum_{\sigma_I = 0}^ {{\cal N}_I -1} \, 
\exp \left( 2 \pi i \sum_j m_j \frac{\sum_{\alpha} 
 \sigma_{\alpha} m_j^{\alpha}} {2k_j + 4 }\right). }
Introducing P-matrices precisely as in \BlumenhagenSU, we finally arrive at the following form of 
the extended Klein bottle amplitude
\eqn\kbf{\eqalign{
\tilde{K}^B &=  \frac{2^5 \prod_j \sqrt {k_j +2}}{2^{\frac{3r}{2} } \,K  \,
\prod_{\alpha} {\cal N}_{\alpha} }\,  \int_0^{\infty} dl \,\frac{1}{\eta(2il)^2}\,  
{\sum_{\lambda',\mu'}}^{ev}\,  \sum_{\nu_0 = 0}^{K-1} \,
\sum_{\nu_1, \ldots, \nu_r=0}^{1} \,
\sum_{\epsilon_1, \ldots, \epsilon_r=0}^{1}\, 
\sum_{\sigma_1= 0}^{{\cal N}_1 -1} \ldots \sum_{\sigma_I= 0}^{{\cal N}_I -1} \,\,
(-1)^{\nu_0} \cr
&\delta^{(4)}_{s'_0, 2 + \nu_0 + 2 \sum \nu_j }\,\,
\prod_{j=1}^r  \bigg( \frac{P^2_{l'_j, \epsilon_j \, k_j}}{S_{l'_j,0}} \,\,
\delta^{(2k_j + 4)}_{m'_j, \nu_0 + (1- \epsilon_j)(k_j + 2) +
\sum_{\alpha} \sigma_{\alpha} m^{\alpha}_j } 
\delta^{(4)}_{s'_j, \nu_0 +2\nu_j + 2(1-\epsilon_j)} \bigg)\,\,
\chi^{\lambda'}_{\mu'}(2il). }}
Consequently, the Ishibashi expansion of the crosscap states extracted from the 
Klein bottle is modified only by the analogous projections on integer monodromy:
\eqn\kbg{\eqalign{
\big| C \big>_B &= \frac{1}{\kappa^B_c}
{\sum_{\lambda',\mu'}}^{ev}\, 
\sum_{\nu_0 = 0}^{K-1} \,
\sum_{\nu_1, \ldots, \nu_r=0}^{1} \,
\sum_{\epsilon_1, \ldots, \epsilon_r=0}^{1}\, 
\sum_{\sigma_1= 0}^{{\cal N}_1 -1} \ldots \sum_{\sigma_I= 0}^{{\cal N}_I -1} \,\,
\Sigma (\lambda', \mu', \nu_0, \nu_j, \epsilon_j, \sigma_{\alpha} m_j^{\alpha}) \cr
&\delta^{(4)}_{s'_0, 2 + \nu_0 + 2 \sum \nu_j }\,\,
\prod_{j=1}^r  \bigg( \frac{P_{l'_j, \epsilon_j \, k_j}}{\sqrt{S_{l'_j,0}}} \,\,
\delta^{(2k_j + 4)}_{m'_j, \nu_0 + (1- \epsilon_j)(k_j + 2) +
\sum_{\alpha} \sigma_{\alpha} m^{\alpha}_j } \,\,
\delta^{(4)}_{s'_j, \nu_0 +2\nu_j + 2(1-\epsilon_j)} \bigg)\,\,
\big|{\lambda'},{\mu'}\big>\big>_c, }}
where again we need to extract the correct sign from the consistent S-transformation of 
the M\"obius-amplitude and the normalization factor is
altered by the prefactors of the Lagrange multipliers as
\eqn\kbh{
(\frac{1}{\kappa^B_c})^2  = \frac { 2^5 \, \prod_{j=1}^r 
\sqrt {k_j +2}}{2^{\frac{3r}{2} } \,K  \,\prod_{\alpha} {\cal N}_{\alpha} }. }
As is clear from the discussion of the A-type Klein bottle, the second 
$\delta$-function above ensures that only those Ishibashi states contribute
in $\tilde{K}^B $ which couple to their charge conjugate in $Z_D$.

\appendix{B}{P-transformation of the M\"obius amplitude}

Starting from
\eqn\appa{
M^{\lambda}_{\mu} = \sum_{\nu_0 = 0}^{ \frac{K}{2} } \, \sum_{ \nu_1, \ldots, \nu_r = 0}^{1} 
\sum_{\tau_1=0}^{{\cal N}_1} \ldots \sum_{\tau_{I'}=0}^{{\cal N}_{I'}} \, (-1)^{[h^{\lambda}_{\mu} (\nu_0, \nu_j, 
\tau_{\alpha}) - h^{\lambda}_{\mu} ]} \,\,
\hat{\chi}^{\lambda}_{\mu + 2 \nu_0 \beta_0 + \sum \nu_j \beta_j +  
\sum_{\alpha=1}^{I'} 2 \tau_{\alpha} m_j^{\alpha} } (it + \frac{1}{2}),}
we perform a P-transformation on the hatted characters by means of the P-matrices as 
given in \BlumenhagenSU.

Collecting the various contributions carefully, we find that we need to evaluate the sum
\eqn\appb{\eqalign{
&\sum_{\nu_0 = 0}^{\frac{K}{2} -1} e^{ 2 \pi i \, \nu_0 \left( 1- 
\sum_{\alpha} 2 \tau_{\alpha} 
Q^{U(1)}(J_{\alpha}) -{s'_0\over 2} + \sum_j \frac{m'_j}{k_j +2} -
\sum_j \frac{s'_j}{2}\right)}
\sum_{\nu_1, \ldots, \nu_j}^{1} 
\left(\prod_{k<l} (-1)^{\nu_k \nu_l} \right) \cr  
&(-1)^{\sum_l \nu_l \left( \frac{s_0 + s_l}{2} + 1 - \frac{s'_l + s'_0}{2} + \epsilon_l
\right) }  
\sum_{\tau_1 = 0}^{{\cal N}_1} \ldots \sum_{\tau_{I'}= 0}^{{\cal N}_{I'}} \,\, \bigg 
( \prod_{\alpha < \beta} ( -1)^{ (2\tau_{\alpha}) \,\,
Q_{\alpha \beta} (2\tau_{\beta})}\bigg) \,\,
e^{ i \pi \sum_{\alpha = 1}^{I'} \tau_{\alpha} x_{\alpha}}, }}
where $ Q^{U(1)}(J_{\alpha})$ denotes the U(1)- charge of $J_{\alpha}$ and 
\eqn\appc{
x_{\alpha} = h(J_{\alpha}) - Q^{(\alpha)}_{\lambda, \mu} + Q^{(\alpha)}_{\lambda', \mu'}
 + \sum_j \epsilon_j \frac{m_j^{\alpha}}{2} }
The sum over $\nu_0$ yields the constraint
\eqn\appd{
Q^{U(1)}_{\lambda', \mu'} - 2 \sum_{\alpha} \tau_{\alpha} Q^{U(1)}(J_{\alpha}) +1 = 0 
\quad {\rm mod} \  1, }
which incorporates  the Gepner projection on odd integer total U(1)-charge; 
the additional Gepner constraints are seen to
be encoded in the various $\delta$-functions appearing in the final expression.

The sum over $\nu_j$ is performed precisely as in \BlumenhagenSU, leading to
\eqn\appe{
(-1)^s \, 2 ^{\frac{r+1}{2}} \, \, \prod_{k<l} (-1)^{\eta_k \eta_l} \,
 \delta^{(2)} _{{\sum_j} \eta_j, 0 } }
with 
\eqn\appf{
\eta_j = \frac{s_0 +s_j}{2} - \frac{s'_0 +s'_j}{2} + \epsilon_j +1. }

As for the evaluation of the sum involving the $\tau_{\alpha}$, we note that due to mutual 
locality and in particular the factors of $2$ appearing
in front of the $\tau_{\alpha}$, the quadratic part equals $1$, and we find immediately the 
desired constraint
\eqn\appf{
Q^{(\alpha)}_{\lambda', \mu'} = 0 \quad {\rm mod} \ 1,  }
taking into account that all other terms in $x_{\alpha}$ are integral anyways. 
Collecting the other trivial factors from the P-matrices, we are lead to the result given 
in the text.

\appendix{C}{$L_j$ quantum numbers of the boundary states}

This is the list of labels for the boundary states in the model
derived from the $(3)^5$ Gepner model:
\eqn\lls{\eqalign{ (L_1,L_2,L_3,L_4,L_5)\in\bigl\{&
(0, 0, 0, 0, 0),(2, 0, 0, 0, 0),(0, 2, 0, 0, 0),(2, 2, 0, 0, 0),(0, 0, 2, 0, 0),\cr
&(2, 0, 2, 0, 0),(0, 2, 2, 0, 0),(2, 2, 2, 0, 0),(0, 0, 0, 2, 0),(2, 0, 0, 2, 0),\cr
&(0, 2, 0, 2, 0),(2, 2, 0, 2, 0),(0, 0, 2, 2, 0),(2, 0, 2, 2, 0),(0, 2, 2, 2, 0),\cr
&(2, 2, 2, 2, 0),(0, 0, 0, 0, 2),(2, 0, 0, 0, 2),(0, 2, 0, 0, 2),(2, 2, 0, 0, 2),\cr
&(0, 0, 2, 0, 2),(2, 0, 2, 0, 2),(0, 2, 2, 0, 2),(2, 2, 2, 0, 2),(0, 0, 0, 2, 2),\cr
&(2, 0, 0, 2, 2),(0, 2, 0, 2, 2),(2, 2, 0, 2, 2),(0, 0, 2, 2, 2),(2, 0, 2, 2, 2),\cr
&(0, 2, 2, 2, 2),(2, 2, 2, 2, 2)\bigr\} }}

And here we list the labels for the boundary states in the model
derived from the $(1)^2\, (7)^3$ Gepner model:
\eqn\llt{\eqalign{ (L_1,L_2,L_3,L_4,L_5)\in\bigl\{&
(0, 0, 0, 0, 0), (0, 0, 2, 0, 0), (0, 0, 4, 0, 0), (0, 0, 6, 0, 0), (0, 0, 0, 2, 0), \cr
&(0, 0, 2, 2, 0), (0, 0, 4, 2, 0), (0, 0, 6, 2, 0), (0, 0, 0, 4, 0), (0, 0, 2, 4, 0), \cr
&(0, 0, 4, 4, 0), (0, 0, 6, 4, 0), (0, 0, 0, 6, 0), (0, 0, 2, 6, 0), (0, 0, 4, 6, 0), \cr
&(0, 0, 6, 6, 0), (0, 0, 0, 0, 2), (0, 0, 2, 0, 2), (0, 0, 4, 0, 2), (0, 0, 6, 0, 2),\cr 
&(0, 0, 0, 2, 2), (0, 0, 2, 2, 2), (0, 0, 4, 2, 2), (0, 0, 6, 2, 2), (0, 0, 0, 4, 2), \cr
&(0, 0, 2, 4, 2), (0, 0, 4, 4, 2), (0, 0, 6, 4, 2), (0, 0, 0, 6, 2), (0, 0, 2, 6, 2), \cr
&(0, 0, 4, 6, 2), (0, 0, 6, 6, 2), (0, 0, 0, 0, 4), (0, 0, 2, 0, 4), (0, 0, 4, 0, 4), \cr
&(0, 0, 6, 0, 4), (0, 0, 0, 2, 4), (0, 0, 2, 2, 4), (0, 0, 4, 2, 4), (0, 0, 6, 2, 4),\cr 
&(0, 0, 0, 4, 4), (0, 0, 2, 4, 4), (0, 0, 4, 4, 4), (0, 0, 6, 4, 4), (0, 0, 0, 6, 4), \cr
&(0, 0, 2, 6, 4), (0, 0, 4, 6, 4), (0, 0, 6, 6, 4), (0, 0, 0, 0, 6), (0, 0, 2, 0, 6), \cr
&(0, 0, 4, 0, 6), (0, 0, 6, 0, 6), (0, 0, 0, 2, 6), (0, 0, 2, 2, 6), (0, 0, 4, 2, 6), \cr
&(0, 0, 6, 2, 6), (0, 0, 0, 4, 6), (0, 0, 2, 4, 6), (0, 0, 4, 4, 6), (0, 0, 6, 4, 6),\cr 
& (0, 0, 0, 6, 6), (0, 0, 2, 6, 6), (0, 0, 4, 6, 6), (0, 0, 6, 6, 6)\bigr\} }}

\vfill\eject
\listrefs

\bye
\end